\documentclass[12pt]{article}

\usepackage{graphicx,a4,amsmath,amssymb}
\usepackage{latexsym,citesort}

\newcommand{\be}{\begin{equation}}
\newcommand{\ee}{\end{equation}}
\newcommand{\bea}{\begin{eqnarray}}
\newcommand{\eea}{\end{eqnarray}}
\newcommand{\eq}{\begin{eqnarray}}
\newcommand{\en}{\end{eqnarray}}
\newcommand{\bc}{\begin{center}}
\newcommand{\ec}{\end{center}}

\newcommand{\nnnl}{\nonumber \\}
\newcommand{\fs}{\,.}
\newcommand{\co}{\,,}
\newcommand{\scs}{\, , \,}

\newcommand{\<}{\langle}
\renewcommand{\>}{\rangle}

\newcommand{\e}{\epsilon}

\newcommand{\GeV}{\,\mbox{GeV}}

\newcommand{\nn}{\nonumber\\}

\newcommand{\mpi}{M_{\pi}}
\newcommand{\mpn}{M_{\pi^0}}
\newcommand{\mpc}{M_\pi}
\newcommand{\mk}{M_K}
\newcommand{\mkl}{M_{K_{\! L}}}

\newcommand{\kl}{K_L}
\newcommand{\phin}{\Phi_0}
\newcommand{\phip}{\Phi_+}
\newcommand{\phim}{\Phi_-}

\newcommand{\abs}[1]{\left|#1\right|}

\newcommand{\Order}{\mathcal{O}}
\newcommand{\M}{\mathcal{M}}

\newcommand{\Li}[1]{\mathrm{Li}_2\left(#1\right)}
\newcommand{\Ec}{E_{\mathrm{cut}}}
\newcommand{\Knn}{(0,0,0)}
\newcommand{\Knc}{(+,-,0)}
\newcommand{\Kcn}{(0,0,+)}
\newcommand{\Kcc}{(+,+,-)}
\newcommand{\Emaxn}{E^{n}_{\mathrm{max}}}
\newcommand{\Emaxp}{E_{\mathrm{max}}}
\newcommand{\Ekin}{E_{\mathrm{kin}}}
\newcommand{\Ekinn}{E_{\mathrm{kin}}^n}

\newcommand{\JC}{J_{+-C}}

\begin{document}

\thispagestyle{empty}

\begin{flushright}
{\footnotesize HISKP--TH--08/10}
\end{flushright}

\vspace{3cm}
\bc{\Large{\bf Radiative corrections in \boldmath{$K\to 3\pi$} decays}}

\vspace{1cm}

M.~Bissegger$^{\,a}$,
A.~Fuhrer$^{\,a,\,b}$,
J.~Gasser$^{\, a}$,
B.~Kubis$^{\, c}$,  
A.~Rusetsky$^{\,c}$

\vspace{2em}

\small
\begin{tabular}{c}
$^a\,$Institute for Theoretical Physics, University of Bern\\   
Sidlerstr. 5, CH-3012 Bern, Switzerland\\[1mm]
$^b\,$Physics Department, University of California, San Diego \\
9500 Gilman Drive, La Jolla, CA 92093-0319, USA\\[1mm] 
$^c\,$Helmholtz-Institut f\"ur Strahlen- und Kernphysik, Universit\"at Bonn\\
Nussallee~14--16, D-53115 Bonn, Germany\\
\end{tabular} 

\ec

\vspace{1cm}

{\abstract{We investigate radiative corrections to $K\to 3\pi$ decays. In particular, we extend the 
non-relativistic framework developed recently to include real and virtual photons and show that, 
in a well-defined power counting scheme, 
the results reproduce corrections obtained in the relativistic calculation.
Real photons are included exactly, beyond the soft-photon approximation, 
and we compare the result with the latter.
The singularities generated by pionium near threshold are investigated, and a region is identified 
where standard perturbation theory in the fine structure constant $\alpha$
may be applied.
 We expect that the formulae provided allow one to extract  S-wave 
 $\pi\pi$ scattering lengths from the cusp effect in these decays with high precision.}}

\vskip1cm

{\footnotesize{\begin{tabular}{ll}
{\bf{Pacs:}}$\!\!\!\!$& 11.30.Rd, 
                        13.20.Eb, 13.40.Ks, 13.75.Lb \\
{\bf{Keywords:}}$\!\!\!\!$& Chiral symmetries, 
       Decays of $K$-mesons, Electromagnetic corrections \\ & to strong- and weak-interaction processes, 
       Meson-meson interactions
\end{tabular}}
}
\clearpage

\tableofcontents

\section{Introduction}

The investigation of the so-called cusp effect  in $K \to 3\pi$
decays~\cite{wigner,fonda,Cabibbo:2004gq} has become a fully competitive method
for the extraction of  S-wave $\pi\pi$ scattering lengths 
from experimental data.
 The theoretical basis for the cusp analysis was provided 
in Refs.~\cite{Cabibbo:2004gq,Cabibbo:2005ez,Gamiz,CGKR,KlongLetter} 
(see also Ref.~\cite{MMS}), whereas  experimental 
results were reported  in Refs.~\cite{Batley,dilellakaon07,MadigozhinCapri,NA48Homepage,KTeV}. 
It turns out that 
the accuracy of the experimental value for the 
difference $a_0-a_2$  of $\pi\pi$ scattering lengths\footnote{We use the 
notation $a_I$ for the (dimensionless)  S-wave scattering 
lengths with isospin $I$.}
 is  mainly limited by  two shortcomings in present theoretical 
descriptions~\cite{Cabibbo:2004gq,Cabibbo:2005ez,Gamiz,CGKR,KlongLetter} 
of the decay amplitudes:  missing contributions  from  $n\geq 3$ loops, and missing 
radiative corrections.
 This article is devoted to an evaluation of the latter. 
 Leaving aside for the moment the contributions from higher loops, 
we  expect that   $K\to 3\pi$ decays, with radiative corrections included, 
and combined with the information gained from $K_{e4}$
decays~\cite{Rosselet,Pislak:2003sv,BatleyKe4,blochFlavia}
 and the pionium lifetime~\cite{dirac1,dirac2}, have the potential to
test the very precise theoretical prediction
of the scattering lengths~\cite{Colangelo:2000jc,Colangelo:2001df}.

We briefly  describe the framework in which we perform the calculation. 
Generically,  we provide 
algebraic expressions for the decay spectra, including the effects of real and virtual photons,  in a 
coherent  approximation scheme which  respects the strictures of unitarity and 
analyticity. The pertinent  expressions for the decay spectra
contain several free parameters, to 
be adjusted such that the experimental distributions are reproduced.  Two of these 
parameters are the S-wave $\pi\pi$  scattering lengths $a_{0,2}$ we are after. 

The cusp effect in both $K^+ \to \pi^0\pi^0\pi^+$ and $K_L \to 3\pi^0$ 
is seen in the decay spectrum
with respect to the invariant mass squared $s_3$ of  $\pi^0\pi^0$ pairs.  
In order to obtain an infrared-finite quantity, one has to take the 
radiation of an additional final-state photon into account as soon as electromagnetic
corrections are included,
\be
\frac{d\Gamma}{ds_3} \biggr|_{E_\gamma<\Emaxp} = 
\frac{d\Gamma(K\to 3\pi)}{ds_3} + \frac{d\Gamma(K\to 3\pi\gamma)}{ds_3} \biggr|_{E_\gamma<\Emaxp} 
+\Order(\alpha^2)\fs\label{eq:decayspect}
\ee
Here, $\alpha=1/137.036$ denotes the fine structure constant, whereas the maximal energy of 
photons not resolved explicitly
in the experiment is denoted  by $\Emaxp$ [measured in the kaon  rest frame]. 
The mentioned approximation scheme is achieved by performing  the calculations
 of the decay spectra in the 
framework of non-relativistic effective field theory (EFT)~\cite{CGKR,KlongLetter},
 extended in this article to include photons.  
 Non-relativistic EFT produces a correlated expansion in $\pi\pi$ scattering
lengths $a$ and in the non-relativistic momentum parameter $\epsilon$, chosen such
that the pion 3-momenta in the final state are counted according to $|\mathbf{p}| = \Order(\e)$.
This expansion is supplemented here with a further expansion parameter $e^2=4\pi\alpha$.

In the following, we will present the photonic corrections  to the  decay spectra 
 up-to-and-including terms of order $e^2a^0\e^4,e^2a^1\e^2$ for the two ``main modes''
$K_L \to 3\pi^0$ 
and $K^+ \to \pi^0\pi^0\pi^+$ that show
a cusp behavior in the invariant $\pi^0\pi^0$ mass spectrum inside the physical region. On the other hand,
in the two ``auxiliary modes'' $K_L \to \pi^+\pi^-\pi^0$ and 
$K^+ \to \pi^+\pi^+\pi^-$, 
 we restrict the accuracy  up-to-and-including terms of order
$e^2a^0\e^4$. Furthermore, we confine ourselves to the so-called
``soft-photon approximation'' in the fully charged channel.
[The framework is set up
in such a manner that the pion masses are not affected by contributions from virtual photons,
as a result of which it is fully consistent to set all pion masses to their physical values
from the very beginning. No expansion in $e^2$ is performed in the pion mass difference.]

Our results can then be represented in the form 
\be
\frac{d\Gamma}{ds_3} \biggr|_{E_\gamma<\Emaxp} = 
\Omega(s_3,\Emaxp) \frac{d\Gamma^{\rm int}}{ds_3} ~
\ee
for the three channels involving at least one neutral pion in the final state, 
and 
\be
d\Gamma \bigr|_{E_\gamma<\Emaxp} = 
\Omega_{++-}(s_1,s_2,s_3,\Emaxp) d\Gamma^{\rm int}  
\ee
for $K^+ \to \pi^+\pi^+\pi^-$,
where both the decay spectrum $d\Gamma^{\rm int}/ds_3$ 
and the differential decay width $d\Gamma^{\rm int}$
are given in terms of a squared matrix element $|\M^{\rm int}|^2$.  
Here, $\Omega(s_3,\Emaxp)$ denotes
a channel-dependent correction factor which is due to real and virtual photons hooked
to charged external legs. All other photonic corrections
 are collected
in the modified matrix element $\M^{\rm int}$.  We will provide both,  the correction factor
and $\M^{\rm int}$, for all four decay channels.

Our work is not the first one to consider radiative
corrections to $K\to 3\pi$ decays -- recent examples include 
Refs.~\cite{nehme,B1,B2,isidorirad,tarasovk3piA,tarasovk3piB}. 
Whereas Isidori's article~\cite{isidorirad} fits
into the framework considered here,  the calculations of Refs.~\cite{nehme,B1,B2}
are of a wider scope and not  useful to extract  $\pi\pi$ 
scattering lengths with high precision by use of  the approach proposed in Ref.~\cite{Cabibbo:2004gq}, 
a framework  also adapted here.  
We refer the interested reader to Sect.~\ref{sec:comparison} 
for a more detailed comparison of our calculation with these works.

The article is composed as follows.
We start out with a brief recapitulation of the non-relativistic EFT framework
without photons in Sect.~\ref{sec:NREFTnoPhotons}.  
The inclusion of photons on the basis of non-relativistic effective Lagrangians
is discussed in Sect.~\ref{sec:Photons}.
Section~\ref{sec:PhotonLoops} treats the essential photon loop diagrams
for the analysis at hand, showing how they can be calculated in non-relativistic EFT, 
from which we derive general power counting rules in Sect.~\ref{sec:powercounting}.
In Sect.~\ref{sec:pionium} we discuss the special role of pionium for 
the cusp region.  Results for the correction factors $\Omega(s_3,\Emaxp)$
and amplitudes $\M^{\rm int}$ are given in Sect.~\ref{sec:results}.
This section contains the main results of our investigation.
We discuss the various sources of theoretical uncertainties for the determination
of the $\pi\pi$ scattering lengths $a_0-a_2$ in Sect.~\ref{sec:accuracy}.
In Sect.~\ref{sec:comparison}, we compare our work to other articles dealing
with radiative corrections in $K\to 3\pi$ decays.
Finally, we close with a summary.
Some background material is relegated to the appendices:
Appendix~\ref{app:notation} collects our notation and explains the necessary kinematics,
Appendices~\ref{app:radcorr}--\ref{app:loops}
contain the calculation of radiative corrections in a relativistic theory.
Appendix~\ref{app:infra-ultra} is
a sample calculation clarifying the interplay between ultraviolet and infrared
divergences in the threshold expansion,
and a discussion of the specific symmetry properties
of the $K_L \to 3\pi^0$ phase space can be found in Appendix~\ref{app:dalitzkl3pi0}.

\begin{sloppypar}
\section{Non-relativistic effective theory in the absence of photons}\label{sec:NREFTnoPhotons}
\end{sloppypar}

We now describe the covariant non-relativistic framework that will be invoked later on.
In this section, we recall the method in the case where electromagnetic corrections are 
discarded~\cite{CGKR,KlongLetter}. 
The inclusion of photons will be discussed in the following sections.

\bigskip

\subsection{Counting rules}

The complete Lagrangian of the effective theory is ${\mathcal L}_K
+{\mathcal L}_{\pi\pi}$, where ${\mathcal L}_K$ contains 
$K^+\to 3\pi$, $\kl\to 3\pi$ vertices,  and ${\mathcal L}_{\pi\pi}$ describes 
elastic $\pi\pi$ scattering in the final state. 
We omit 6-pion couplings, see later.
The interactions are ordered according to specific
counting rules~\cite{CGKR,KlongLetter}. A formal small parameter
$\epsilon$ is introduced, and the various quantities are counted as follows.
\begin{itemize}
\item
The masses $M_K,M_{K_L},M_\pi,M_{\pi^0}$ are counted as $\Order(1)$.
\item
All three-momenta (spatial derivatives) are counted as $\Order(\epsilon)$.
\item
Kinetic energies are counted as $\Order(\epsilon^2)$.
\end{itemize}
As a result of this, one has to further count
\begin{itemize}
\item
$M_K-3M_\pi=\Order(\epsilon^2)$, $M_{K_L}-3M_\pi=\Order(\epsilon^2)$.
\item
$\mpc - \mpn =\Order(\epsilon^2)$.
\end{itemize}
Here, $M_K,M_{K_L},M_\pi,M_{\pi^0}$ denote the masses of $K^+,K_L,\pi^+,\pi^0$,
respectively.

We shall treat the pion-pion interaction perturbatively, owing
to the smallness of the pion-pion scattering lengths. For bookkeeping reasons,
we introduce an additional
formal small parameter $a$, which stands for the pion-pion scattering length.
All four-pion couplings are regarded as quantities of order
$a$. At the end, the amplitudes are given in the form of a double expansion
in $\epsilon$ and $a$. This expansion is correlated, since adding one more
pion loop with an additional four-pion vertex also increases the power of $\epsilon$ by one.

In the following, we provide the Lagrangians necessary to carry out the
calculation of the amplitudes for $K^+\to 3\pi$, $\kl\to 3\pi$.

\subsection{The Lagrangians}\label{sec:pipi}

\begin{sloppypar}
We start with the $\pi\pi$ interaction and consider the
following five physical channels in $\pi^a\pi^b\to\pi^c\pi^d$: $(ab;cd)=$
(1)~$(00;00)$, (2)~$(+0;+0)$, (3)~$(+-;00)$, (4)~$(+-;+-)$, (5)~$(++;++)$.
 [We omit the channel $\pi^-\pi^0\to\pi^-\pi^0$, because this amplitude is 
identical to $\pi^+\pi^0\to\pi^+\pi^0$ by charge invariance.] The Lagrangian takes the form
\be\label{eq:l_pipi}
{\mathcal L}_{\pi\pi}=2\sum_\pm\Phi_\pm^\dagger W_\pm\bigl(i\partial_t-
W_\pm\bigr)\Phi_\pm 
+2\Phi_0^\dagger W_0\bigl(i\partial_t- W_0\bigr)\Phi_0
+\sum_{i=1}^5{\mathcal L}_{i}\, ,
\ee
where $\Phi_m$, $m=\pm,\,0$ are non-relativistic pion field operators, 
 and $W_m=\sqrt{M_{\pi^m}^2-\triangle}$,   with
$\triangle$ the Laplacian. 
At leading order in the non-relativistic expansion, one simply finds
\be \label{eq:l_1-5}
{\mathcal L}_{i} = x_iC_i
\bigl(\Phi_c^\dagger\Phi_d^\dagger\Phi_a\Phi_b+h.c.\bigr)
+\ldots~,
\ee
where the ellipsis stands for terms of order $\epsilon^2$ and higher. 
Higher-order terms are explicitly given in Refs.~\cite{CGKR,KlongLetter}.
The low-energy constants $C_i$ 
are matched to the physical threshold amplitudes below in Sect.~\ref{subsec:matching}.
To simplify the resulting expressions, we have furthermore 
introduced the combinatorial factors $x_1=x_5=1/4$, $x_2=x_3=x_4=1$.
Finally,  we note that we omit local 6-pion couplings.
Their contribution to the $K\to 3\pi$ amplitudes is purely imaginary in
the non-relativistic framework, and of order $\epsilon^4$.
\end{sloppypar}

Next, we consider the non-relativistic Lagrangians that describe $K\to 3\pi$
decays in different channels. The following notation is useful:
\newcommand{\Pp}{\mathcal{P}}
\begin{align}
(K^\dagger)_\mu &= ({\cal P}^\dagger)_\mu K^\dagger \, , & \!\!
({\cal P}^\dagger)_\mu &= (W_K,i\nabla)\, , & \!\!
(K_L^\dagger)_\mu &= (\bar {\cal P}^\dagger)_\mu K_L^\dagger\, , & \!\!
(\bar{\cal P}^\dagger)_\mu &= (\bar W_K,i\nabla)\, , \nonumber \\
\label{eq:Pn}
(\Phi_n)_\mu &= (\Pp_n)_\mu\Phi_n\, , & \!\!
(\Pp_n)_\mu &= (W_n,-i\nabla)\, , & \!\!
(\Phi_n^\dagger)_\mu &= (\Pp^\dagger_n)_\mu\Phi^\dagger_n\, ,& \!\!
(\Pp_n^\dagger)_\mu &= (W_n,i\nabla)\, ,
\end{align}
for $n=a,\,b,\,c,\,d$,
where $K,K_L$ denote the non-relativistic fields for the $K^+,\kl$ mesons,
and $ W_K=\sqrt{M_K^2-\triangle}$, $\bar W_K=\sqrt{M_{K_L}^2-\triangle}$.
The Lagrangians that describe $K^+\to 3\pi$ and $\kl\rightarrow 3\pi$
decays  at tree level are given by
\begin{align}
\label{eq:lag_K}
{\mathcal L}_K &= 2K^\dagger W_K\bigl(i\partial_t- W_K\bigr)K
+\frac{1}{2}\, G_0\,\bigl(K^\dagger\Phi_+\Phi_0^2+h.c.\bigr)
\nonumber\\[2mm]
&\quad + \frac{1}{2}\, G_1\,\biggl\{
\biggl(\frac{(K^\dagger)_\mu(\Phi_+)^\mu}{M_K}
-M_\pi K^\dagger\Phi_+\biggr)
\Phi_0^2+h.c.\biggr\}
\nonumber\\[2mm]
&\quad + \frac{1}{2}\, H_0\,\bigl(K^\dagger\Phi_-\Phi_+^2+h.c.\bigr)
\nonumber\\[2mm]
&\quad + \frac{1}{2}\, H_1\,\biggl\{
\biggl(\frac{(K^\dagger)_\mu(\Phi_-)^\mu}{M_K}
-M_\pi K^\dagger\Phi_-\biggr) \Phi_+^2+h.c.\biggr\}
+\ldots\, , \\[3mm] 
\label{eq:lag_KL}
\bar {\mathcal L}_K &= 2K_L^\dagger \bar W_K\bigl(i\partial_t- \bar W_K\bigr)K_L
 +L_0\bigl(K_L^\dagger\phin\phip\phim+h.c.\bigr)
\nonumber\\[2mm]
&\quad + L_1\biggl\{
\biggl(\frac{(K_L^\dagger)_\mu(\Phi_0)^\mu}{M_{K_L}}
-M_{\pi^0} K_L^\dagger\Phi_0\biggr)
\Phi_+\Phi_-+h.c.\biggr\}
\nonumber\\[2mm]
&\quad + \frac{1}{6}K_0\bigl(K_L^\dagger\phin^3+h.c.\bigr)
+ \ldots\,,
\end{align}
where the ellipsis stands for the
higher-order terms in $\epsilon$. The couplings $G_i,H_i,L_i$, $K_i$ are
assumed to be real. 

\subsection{Loops}

The Lagrangians displayed above generate pion loop contributions, according 
to the 
standard Feynman diagrammatic technique. The 
pion propagator is given by
\eq
S_m(p)=\frac{1}{2w_m({\bf p})}\,\frac{1}{w_m({\bf p})-p^0-i0}\, ,
\en
where $w_m({\bf p})=\sqrt{M_{\pi^m}^2+{\bf p}^2}$. The 
charged and neutral kaon propagators $S_K(p)$, $\bar S_K(p)$
are defined similarly.

However, since the heavy scales
(pion and kaon masses) are explicitly present in the particle propagators,
the simple and straightforward counting rules 
in $\epsilon$, which were introduced at tree level, will be ruined by
loop corrections. As it is well known (see, e.g., Refs.~\cite{Beneke,nrqft7,physrep}),
the counting rules can be restored, imposing additional prescriptions 
(the threshold expansion) built 
on top of the Feynman rules. In our case, these prescriptions state:
\begin{itemize}
\item
For a given Feynman integral, expand all inhomogeneous factors (e.g.\ square roots) 
in the propagators in powers of the inverse pion masses.
\item
Integrate the expanded Feynman integral term by term.
\item
Re-sum the final result to all orders of the inverse pion masses.
\end{itemize}
It should be noted that the framework is more complicated technically (due to 
the presence of the square roots that ensure the correct relativistic 
dispersion law for the particles in the loops) than the 
standard non-relativistic EFT (see, e.g., Ref.~\cite{physrep}). However, the bonus
is that the location of the low-energy singularities in different amplitudes
is Lorentz-covariant. This is an important advantage in the study of 
three-particle final states, where different two-particle sub-systems 
are in general not in the center-of-mass frame. Note also that, without re-summing
the final result, we again arrive at the standard formulation of the 
non-relativistic EFT.

The procedure outlined above has been carried out up to 
two loops~\cite{CGKR,KlongLetter,GKR}. Details of the calculations will be provided
in Ref.~\cite{GKR}. We only give a short summary here.

The one-loop contributions are proportional to the basic integral
\eq\label{eq:Jab}
J_{ab}(s)=\int\frac{d^Dl}{i(2\pi)^D}\, S_a(P-l)S_b(l)
=\frac{i\lambda^{1/2}(s,M_a^2,M_b^2)}{16\pi s}\, ,
\en
where $s=P^2$, and $\lambda(x,y,z)=x^2+y^2+z^2-2xy-2yz-2zx$ denotes 
the triangle function.
We see that,
in contrast to the conventional non-relativistic framework,
the basic integral is explicitly covariant.\footnote{This result can also be 
obtained in the framework developed in Refs.~\cite{Goity,Lehmann}.}

\begin{figure}[t]
\begin{center}
\includegraphics[width=0.8\linewidth]{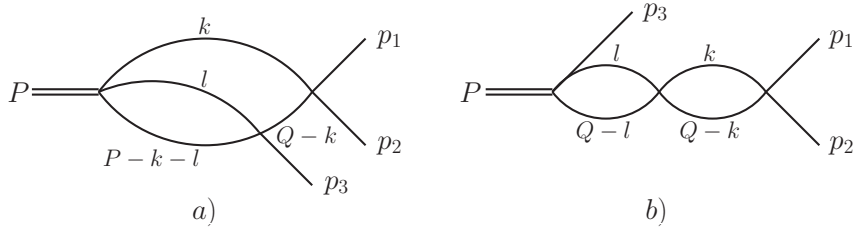}
\end{center}
\caption{Two topologically distinct non-relativistic two-loop graphs 
describing the final-state $\pi\pi$ re-scattering in the decay $K\to 3\pi$, with
$Q^\mu=(p_1+p_2)^\mu$.}
\label{fig:2loop_typ}
\end{figure}

There are two types of two-loop diagrams with 4-pion vertices, shown
in Fig.~\ref{fig:2loop_typ}. At the order we are 
working, it suffices to consider the
diagrams in Fig.~\ref{fig:2loop_typ} with non-derivative couplings only.
In this case, the contributions of 
both diagrams depend only on the variable $s$, where
\be
s=Q^2 ~, \quad  Q^\mu=(p_1+p_2)^\mu ~.
\ee

The non-trivial contribution from Fig.~\ref{fig:2loop_typ}a
is proportional to
\be\label{eq:twoloop_F}
{\mathcal M}(s)
=\int\frac{d^D l}{i(2\pi)^D}\,\frac{d^D k}{i(2\pi)^D}\,
S_a(P-l-k)S_b(l)S_c(k)S_d(Q-k)
\ee
A short discussion of this integral is given  in Ref.~\cite{CGKR}. There, it is shown that one may write
\be\label{eq:mostgeneral}
{\mathcal M}(s)
=F\bigl(M_a,M_b,M_c,M_d;s\bigr)+\ldots ,
\ee
where $F$ is ultraviolet finite and
contains the full non-analytic behavior of the two-loop diagram
in the low-energy domain, whereas the ellipsis denotes terms that amount to a redefinition 
of the tree-level couplings in ${\cal L}_K,\bar {\cal L}_K$ 
and which are therefore dropped.
 A one-dimensional  integral representation for $F$  
is provided in Ref.~\cite{CGKR}.
The relevant integrals can be also performed analytically~\cite{KlongLetter}.

Finally, the diagram in Fig.~\ref{fig:2loop_typ}b is a product of two one-loop 
diagrams given in Eq.~\eqref{eq:Jab}.

\section{Including photons}\label{sec:Photons}

\subsection{The Lagrangian with photons}
\label{sec:PhotonsTreeLevel}

In this and the following sections we discuss the inclusion of photons in the non-relativistic 
framework. Since power counting is more complicated in the presence of
photons, we first consider examples of simple vertices and 
diagrams with photons,
without trying to order them according to the power counting.

The kinetic part of the Lagrangian after minimal substitution takes the form
\begin{align}
{\cal L}_{\rm kin} &= -\frac{1}{4}\,F_{\mu\nu}F^{\mu\nu}
+\sum_\pm\Bigl(i\Phi_\pm^\dagger D_t{\cal W}_\pm\Phi_\pm
-i(D_t{\cal W}_\pm\Phi_\pm)^\dagger\Phi_\pm-2\Phi_\pm^\dagger {\cal W}_\pm^2\Phi_\pm\Bigr)
\nonumber\\
&+ 2\Phi_0^\dagger W_0(i\partial_t-W_0)\Phi_0
+\Bigl(iK^\dagger D_t{\cal W}_K K
-i(D_t{\cal W}_K K)^\dagger K-2K^\dagger {\cal W}_K^2 K\Bigr)
\nonumber\\[2mm]
&+ 2K_L^\dagger \bar W_K (i\partial_t-\bar W_K)K_L\, .\label{eq:PhotonLagr}
\end{align}
Here, $F_{\mu\nu}=\partial_\mu A_\nu - \partial_\nu A_\mu$ denotes the electromagnetic field strength tensor and
\begin{align}
D_t\Phi_\pm &= (\partial_t\mp ieA_0)\Phi_\pm\, , &
D_tK &= (\partial_t- ieA_0)K\, ,
\nonumber\\[2mm]
{\cal W}_\pm &= \sqrt{M_\pi^2-{\bf D}^2}\, , &
{\cal W}_K &= \sqrt{M_K^2-{\bf D}^2}\, ,
\nonumber\\[2mm]
{\bf D}\Phi_\pm &= (\nabla\pm ie{\bf A})\Phi_\pm\, , &
{\bf D}K &= (\nabla + ie{\bf A})K\, .
\end{align}
In an explicit calculation of the tree-level photon-pion and photon-kaon vertices,
the roots ${\cal W}_\pm$, ${\cal W}_K$ are understood to be expanded in the inverse 
of the corresponding mass. After resummation of all terms proportional to
$e$, the relativistic result is explicitly reproduced at tree level,
\eq
\langle\pi^\pm(p')|j_\mu^{\rm em}(0)|\pi^\pm(p)\rangle=\pm e(p'+p)_\mu\, ,~
\langle K^+(p')|j_\mu^{\rm em}(0)|K^+(p)\rangle= e(p'+p)_\mu\, .
\en
$j_\mu^{\rm em}$ denotes the electromagnetic current, calculated 
from the Lagrangian in a standard manner.
Note that the non-relativistic Lagrangian Eq.~\eqref{eq:PhotonLagr} does not contain
a term for pion pair creation by a photon, $\gamma \to \pi^+ \pi^-$,
which would involve a high-energy photon.
This omission eliminates all one-photon reducible graphs from the $K\to 3\pi$ 
amplitudes considered in the following.

\begin{figure}
\begin{center}
\includegraphics[width=2.5cm]{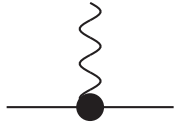}
\end{center}
\caption{Effective-range coupling of the Coulomb photon 
to a charged pion.  The coupling is proportional to the pion charge radius squared.}\label{fig:effrange}
\end{figure}

The spatial derivatives in the four-pion vertex are replaced by the covariant
derivatives in the presence of photons. 
In addition, non-minimal terms, containing the electric ${\bf E}$ and
magnetic ${\bf B}$ fields, will be present.
At lowest order, the non-local coupling of the Coulomb photon field $A_0$
with e.g.\ the pion emerges, see 
Fig.~\ref{fig:effrange}. The pertinent coupling constant is proportional to
the mean charge radius squared of the pion~\cite{physrep}. However, at the order we are
working, this term in the Lagrangian can be eliminated by using the 
equations of
motion and thus it does not affect any physical observable (we refer the
interested reader to Ref.~\cite{physrep} for a detailed discussion of this
issue). It can also be  shown that higher-order non-minimal couplings do
not contribute at the order we are working.
In the following, we discard all non-minimal couplings 
in the Lagrangian from the beginning. 
Further,  we also discard  the contribution arising from 
vacuum polarization generated by an electron-positron loop (see again Ref.~\cite{physrep}).
We have checked that this correction 
is tiny and
amounts to a few percent change of the (already small) Coulomb correction in
the ``auxiliary'' decay modes.

The loop calculations in the non-relativistic theory are most easily done
in the Coulomb gauge. Here, first the time component $A_0$ is removed from
the Lagrangian by using the equations of motion.
This procedure generates the non-local operator
\eq
\triangle^{-1}=-\int\frac{d^3{\bf k}}{(2\pi)^3}\,
\frac{{\rm e}^{-i{\bf k}({\bf x}-{\bf y})}}{{\bf k}^2}
\en
in the Lagrangian. In the following, for ease of understanding, we keep
calling the pertinent diagrams as ``generated by the exchange of Coulomb 
photons.'' The propagator of the transverse photons in the Coulomb gauge
is given by
\eq
D^{ij}(k)=i\int d^4x\,{\rm e}^{ikx}\langle 0|TA^i(x)A^j(0)|0\rangle
=-\frac{1}{k^2}\,\biggl(\delta^{ij}-\frac{k^ik^j}{{\bf k}^2}\biggr)\, .
\en
Below, the virtual photon corrections to the strong amplitudes are calculated
at $\Order(e^2)$ (with the exception of the discussion in Sect.~\ref{sec:pionium}), 
i.e.\ we include one virtual photon at most.
In order to obey counting rules, as in the purely strong case,
one has to perform threshold expansions. In the following section we 
explain how this can be done on the basis of several examples. The counting rules
in the presence of photons are discussed in  full generality 
afterwards in Sect.~\ref{sec:powercounting}.

\subsection{Matching}
\label{subsec:matching}

The effective coupling constants in the Lagrangian can be related to
physical observables in the underlying relativistic field theory. This
procedure goes under the name of {\it matching}. Here, chiral perturbation theory is understood to be 
the underlying theory, which we bear in mind while
performing the matching.

In order to determine the (four-pion) coupling constants in the
non-relativistic theory through matching, one has to calculate the pion-pion
scattering amplitude both in the non-relativistic theory and in the underlying
theory. These amplitudes are then expanded in the pion momenta at
threshold, and the pertinent expansion coefficients are set equal (up to a
given order). One has to distinguish the isospin symmetry limit and the case
with broken isospin symmetry.
 
\subsubsection{Isospin symmetry limit}

In the isospin symmetry limit,
the expansion of the relativistic $\pi\pi$ scattering amplitude at threshold 
reads
\be
{\rm Re}\,\bar T_i(s,t) =
\bar A_i + \Order(\e^2) ~.
\ee 
The bar indicates the quantities in the isospin-symmetric world, 
in which the pion has the mass 
$M_\pi=139.57$~MeV. In terms of the standard  scattering
lengths $a_0$ and $a_2$, one has
\begin{align}
\label{eq:isospin}
3\bar A_1 &= {N(a_0+2a_2)} \co &
2\bar A_2 &= {Na_2} \co &
3\bar A_3 &= {N(a_2-a_0)} \co \nnnl
6\bar A_4 &= {N(2a_0+a_2)} \co &
\bar A_5 &= Na_2\co & N &= 32 \pi \,\,, 
\end{align}
with $a_0=0.220\pm 0.005$, $a_2=-0.0444\pm 0.0010$, $a_0-a_2=0.265\pm 0.004$~\cite{Colangelo:2001df}. 
Still in the isospin symmetry limit, the couplings $C_{i}$ are 
related to these threshold parameters according to 
\be\label{eq:pipimatching1}
 2\bar C_i=\bar A_i\,\,\,, 
\ee
where we have dropped higher-order terms in the threshold parameters.

We wish to note that the matching condition is universal in the
isospin-symmetric case and determines the coupling constants in terms of the
effective-range expansion parameters only. 
There is no explicit reference to the 
Lagrangian of the underlying relativistic theory.

\subsubsection{Isospin-violating case}

The isospin-breaking corrections to the isospin-symmetric result
emerge from electromagnetic effects as well as from the up/down
quark mass difference and have the following general 
form (see, e.g., Refs.~\cite{nrqft7,physrep}):
\eq
C_i=\bar C_i+h_1(m_d-m_u)^2+h_2e^2+\Order(e^4,e^2(m_d-m_u)^2,(m_d-m_u)^4)\, ,
\en
where the coefficients $h_{1,2}$ depend on the quark mass $\hat m=\frac{1}{2}\,(m_u+m_d)$.
Note that the corrections are no more universal, and 
in order to carry out the calculations, 
 one has to use the explicit form of 
the underlying Lagrangian of the relativistic theory. Further, in order to 
perform the matching in the presence of photons, one has first to remove the
infrared-singular piece of the amplitude at threshold. The procedure is 
described in Refs.~\cite{nrqft7,physrep} and will not be repeated here.

At leading order in chiral perturbation theory one finds~\cite{knechturech,nrqft7} 
\be\label{eq:pipimatching2}
2C_{1,2,5}=\bar A_{1,2,5}(1-\eta),\,\, 
2C_{3}=\bar A_3(1+\eta/3),\,\,
2C_{4}=\bar A_4(1+\eta), 
\ee
where $\eta=(\mpc^2-\mpn^2)/\mpc^2=6.5\times 10^{-2}$.

The calculations at one-loop level have been carried out, in particular, 
for the process $\pi^+\pi^-\to\pi^0\pi^0$ that multiplies the cusp in the 
$K\to 3\pi$ decays at lowest order. The final result can be extracted, e.g.,
from  Eqs.~(4.14), (4.28) and (4.29) of Ref.~\cite{nrqft7},
\eq
2C_3&=& -\frac{32\pi}{3} \bigl[a_0-a_2+(0.61\pm 0.16)\times 10^{-2}\bigr]
\nonumber\\[2mm]
&=&\bar A_3\,(1+(2.3\pm 0.6)\times 10^{-2})\, ,
\en
where in the second line we have used $a_0-a_2=0.265$.
The bulk of the total correction is already given by the tree-level result in 
Eq.~\eqref{eq:pipimatching2}, and
most of the uncertainty stems from the so-called ``electromagnetic''
low-energy constants in the $\Order(e^2p^2)$ chiral 
Lagrangian~\cite{knechturech}.
The higher-order corrections to the matching of $C_{1,2,4,5}$ are less important, because these couplings
enter the cusp amplitude at the sub-leading order. In general, one may 
conclude that  isospin-breaking corrections beyond tree 
level -- given in Eq.~\eqref{eq:pipimatching2} -- amount to a systematic uncertainty in the
$C_i$ at the percent level.

\section{Examples of loop calculations with photons}\label{sec:PhotonLoops}

\subsection{Ultraviolet and infrared divergences: general remarks}

The loop diagrams with photons in the non-relativistic EFT 
have both ultraviolet and infrared divergences.
Their treatment is different: whereas  
the ultraviolet divergences are removed by renormalization, the infrared
ones cancel in the total decay rate, including the decay into soft
photons. Further, the non-relativistic EFT  exactly reproduces 
the structure of the
infrared divergences in the underlying relativistic theory. 
On the other hand, the high-energy behavior in these two theories, in general, differs.

Throughout this paper, we tame both ultraviolet and infrared divergences
by dimensional regularization. In addition, the threshold expansion is used
in the Feynman integrals in order to ensure the validity of the counting rules 
in the presence of photons. One has to face, however, the following problem.
Both ultraviolet and infrared divergences show up as poles in the amplitudes 
in the variable $D$ as $D\to 4$. The origin of the singularity (either ultraviolet
or infrared) can be easily identified, and one may attach a subscript to these 
poles, writing $\left(\frac{1}{D-4}\right)_{UV}$ and 
$\left(\frac{1}{D-4}\right)_{IR}$ explicitly.
At the order we are working, it can be
checked that, if threshold expansion is not applied, the infrared poles that emerge in the
non-relativistic EFT, as expected, are identical to those in the pertinent 
graphs of the underlying relativistic theory. 

However, the threshold expansion generally
changes the behavior of the diagrams in the infrared, and the structure of the infrared
poles does not match the underlying theory anymore. In order to see this, we note that
applying threshold expansion and the subsequent use of the no-scale argument in 
dimensional regularization amounts to the replacement
\eq
\left(\frac{1}{D-4}\right)_{UV}-\left(\frac{1}{D-4}\right)_{IR}\to 0
\en
in the pertinent terms. For more details, we refer the reader to
Appendix~\ref{app:infra-ultra}, where we illustrate the issue with the  vertex diagram.

In this paper, we adopt the following simple solution of the problem. 
As becomes clear from the above expression, the coefficient 
of the infrared pole changes by a low-energy ``polynomial'' as a result of the threshold expansion.
For this reason, we do not differentiate between infrared and ultraviolet divergences 
(detaching the subscripts ``UV'' and ``IR'' everywhere) and remove
all remaining divergences by the counterterms of the Lagrangian in a threshold-expanded EFT.
This leads to the correct result for all observables, 
since all infrared divergences  cancel at the end.

\begin{figure}[t]
\begin{center}
\includegraphics[width=0.8\linewidth]{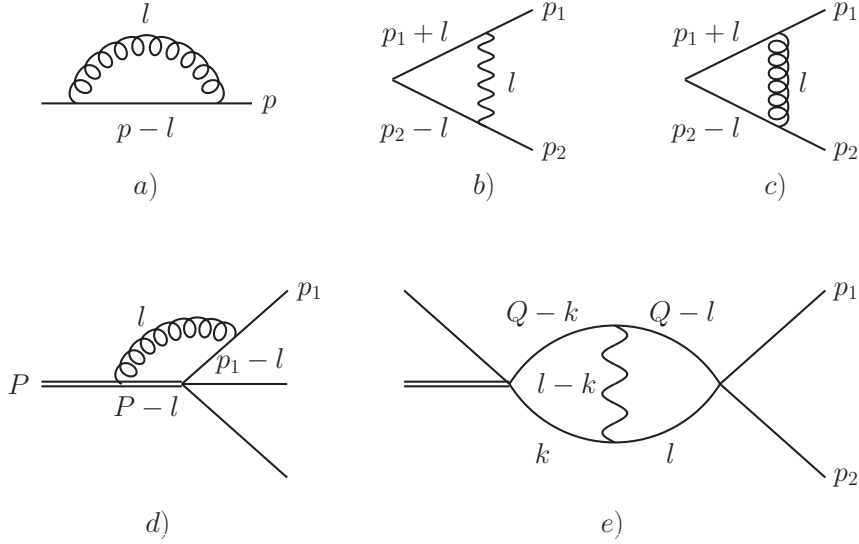}
\end{center}
\caption{Examples of loop graphs with photons in the non-relativistic
EFT.  Double and single full lines denote kaons and pions, respectively, 
wiggly lines stand for Coulomb, curly lines for transverse photons.
a: pion self-energy (transverse photon); b: vertex diagram with Coulomb
photon; c: vertex diagram with transverse photon; d: vertex diagram in crossed 
channel (transverse photon); e: Coulomb photon exchange in the strong loop.}
\label{fig:demo}
\end{figure}

\subsection{Self-energy of the pion}\label{sec:selfenergy}

The self-energy diagram of the charged pion shown in Fig.~\ref{fig:demo}a
is given by the following integral
\eq
\Sigma(p)=\frac{e^2}{2w_\pm({\bf p})}\,\int\frac{d^Dl}{i(2\pi)^D}\,S_\pm(p-l)
\frac{(2p-l)^i(2p-l)^j}{-l^2}
\biggl(\delta^{ij}-\frac{l^il^j}{{\bf l}^2}\biggr)\, .
\en
Only the transverse photons contribute.

In order to calculate $\Sigma(p)$, we perform the Cauchy integral over $l^0$
and further use the identity
\begin{align}
\frac{1}{2w_\pm({\bf p}-{\bf l})}\,\frac{1}{w_\pm({\bf p}-{\bf l})-p^0+|{\bf l}|}
&= -\frac{1}{2w_\pm({\bf p}-{\bf l})}\,\frac{1}{w_\pm({\bf p}-{\bf l})+p^0-|{\bf l}|}
\nonumber\\[2mm]
&\quad + \frac{1}{w_\pm^2({\bf p}-{\bf l})-(p^0-|{\bf l}|)^2}\, .
\end{align}
Since both $w_\pm({\bf p}-{\bf l}),~p^0=\Order(1)$ in the $\epsilon$ counting, the
threshold expansion in the first term generates polynomials in the integration
momenta. Using no-scale arguments, it 
is seen that the integral of the first term vanishes in dimensional 
regularization. 

The denominator in the second term can be written as
$\Omega+2p^0|{\bf l}|-2{\bf p}{\bf l}$, where $\Omega=w_\pm^2({\bf p})-(p^0)^2
=M_\pi^2-p^2=\Order(\epsilon^2)$. In order to expand the denominator, one has 
to establish the power counting rules for the photon 3-momentum ${\bf l}$.
One observes that in this particular diagram  ${\bf l}$
should be assigned the power $\Order(\epsilon^2)$, 
otherwise one arrives at no-scale integrals after expansion in $\e$,
which vanish in  dimensional regularization. Assuming
${\bf l}=\Order(\epsilon^2)$ and expanding this denominator in the last term, which
is of order $\epsilon^3$ (the first two terms are of order $\epsilon^2$),
we find\footnote{Note that here and in the following, the results we show
for various diagrams are always to be understood with the threshold expansion applied.}
\be\label{eq:Sigma}
\Sigma(p)=\frac{e^2{\bf p}^2}{2w_\pm({\bf p})p^0}
\biggl(\frac{\Omega}{2p^0}\biggr)^{d-2}\Bigl(1-\frac{1}{d}\Bigr)
\frac{2\pi^{d/2}}{(2\pi)^d\Gamma(\frac{d}{2})}\,\Gamma(d-1)\Gamma(2-d)
+\ldots \, ,
\ee
where $d=D-1$, and the ellipses stand for higher-order terms in the
$\e$-expansion. In the limit $d\to 3$ we obtain
\begin{align}
\Sigma(p)&= \frac{4e^2{\bf p}^2\Omega}{3w_\pm({\bf p})(p^0)^2}\,
\biggl(\lambda+\frac{1}{16\pi^2}\biggl(\ln\frac{\Omega}{\mu p^0}-\frac{1}{3}\biggr)\biggr)
+ \ldots \, ,
\\[2mm]
 \label{eq:lambda}
\lambda &= \frac{\mu^{D-4}}{16\pi^2}\,\biggl(\frac{1}{D-4}-\frac{1}{2}\,\bigl(\Gamma'(1)+\ln 4\pi+1\bigr)
\biggr)\, ,
\end{align}
which is in perfect agreement with the standard non-relativistic framework
(see e.g.\ Ref.~\cite{nrqft7}).
Here $\mu$ denotes the scale of dimensional regularization. The divergence at $D\to 4$ is 
ultraviolet. However, as mentioned above, in the following we shall not distinguish
between the ultraviolet and infrared poles. Note that, since this 
expansion does not move the location of the singularity in the self-energy
fixed by the relativistic relation $p^2=M_\pi^2$, there is no need to 
eventually re-sum the obtained series.

The expression for the $Z$-factor of the charged pion
can be directly obtained from 
Eq.~\eqref{eq:Sigma} by performing the limit $\Omega\to 0$ at
$d> 3$ (see, e.g., Refs.~\cite{annals,physrep}).  In this manner, one finds
\eq
Z=1 \, .
\en

\subsection{Coulomb vertex}\label{sec:Coulombvertex}

The Coulomb vertex, shown in Fig.~\ref{fig:demo}b, is given by
\begin{align}
V_C &= -e^2\int\frac{d^Dl}{i(2\pi)^D}\, S_\pm(p_1+l)S_\pm(p_2-l)\,
\frac{(w_1+w_1')(w_2+w_2')}{{\bf l}^2}\, ,
\nonumber\\[2mm]
w_1 &= w_\pm({\bf p}_1)\, ,\quad
w_1'=w_\pm({\bf p}_1+{\bf l})\, ,\quad
w_2=w_\pm({\bf p}_2)\, ,\quad
w_2'=w_\pm({\bf p}_2-{\bf l})\, .
\end{align}
Performing the Cauchy integration over $l^0$
and using the threshold expansion, one can demonstrate
that the numerator may be replaced by
\eq
(w_1+w_1')(w_2+w_2')\to 4w_1w_2
-\biggl(\frac{w_1-w_2}{w_1+w_2}\,{\bf Q}+2{\bf q}\biggr)\,{\bf l}-{\bf l}^2\, ,
\en
with ${\bf Q}={\bf p}_1+{\bf p}_2$ and ${\bf q}=({\bf p}_1-{\bf p}_2)/2$. 
Further, it can be 
checked that the energy denominator that emerges after integration over 
$l^0$ can be rewritten as
\begin{align}
\frac{1}{2w_1'2w_2'}\,\frac{1}{w_1'+w_2'-E} &= \frac{1}{2E}\,\frac{1}
{{\bf l}^2+2{\bf b}{\bf l}-({\bf Q}{\bf l})^2/E^2
-\lambda(E^2,w_1^2,w_2^2)/4E^2}+\ldots\, ,
\nonumber\\[2mm]
E &= w_1+w_2\, , \qquad
{\bf b} = 
{\bf q}-\frac{{\bf Q}({\bf Q}{\bf q})}{E^2}\, .
\end{align}
Using the substitution
\eq
{\bf l}\to {\bf l}+\frac{{\bf Q}({\bf Q}{\bf l})}{\sqrt{s}(E+\sqrt{s})}
-{\bf b}-\frac{{\bf Q}({\bf Q}{\bf b})}{s}\, ,\quad\quad
s=E^2-{\bf Q}^2 = (p_1+p_2)^2 \, ,
\en
we may rewrite the integral as
\eq\label{eq:VC-nonrel}
V_C=-\frac{e^2}{2\sqrt{s}}\,\int\frac{d^dl}{(2\pi)^d}\,
\frac{1}{{\bf l}^2-q_0^2}\,\frac{4M_\pi^2}{({\bf l}-q_0{\bf q}/|{\bf q}|)^2} + \ldots\, ,\quad\quad
q_0^2=\frac{s}{4}-M_\pi^2\, ,
\en
where the ellipses again represent higher-order terms in $\e$. 
At lowest order in $\epsilon$ we obtain (cf.\  e.g.\ Refs.~\cite{nrqft7,physrep})
\eq
V_C=\frac{2\alpha M_\pi }{\sqrt{s}}\,
\biggl(-\frac{\pi M_\pi}{4q_0}-i \theta_c(q_0)+\Order(d-3)\biggr)+\ldots\, , \quad q_0^2 > 0 \, ,
\label{eq:VC}
\en
where $\alpha$ denotes the fine structure constant,
\eq
\theta_c(q_0)=\frac{M_\pi}{2q_0}\biggl( 16\pi^2\lambda
+\frac{1}{2}\biggl(\ln\frac{4q_0^2}{\mu^2}+1\biggr)\biggr) + \ldots \label{eq:thetac}
\en
is the (infrared-divergent) Coulomb phase, and the ellipses stand for 
higher-order terms in the expansion in $\epsilon$. Since these terms do not 
change the position of the branch point in the diagram at $s=4M_\pi^2$,
there is no need to re-sum these terms.

{}From Eq.~\eqref{eq:VC-nonrel} we may observe that the counting of
the photon momentum ${\bf l}=\Order(\epsilon)$ is different from the case 
of the self-energy diagram. 

\subsection{Transverse vertex}\label{sec:transvert}

The vertex with the transverse photon, 
which is shown in Fig.~\ref{fig:demo}c, is 
suppressed by a factor $\epsilon^2$ as compared to the Coulomb vertex
due to the derivative coupling of the transverse photons to pions.
It is given by the following Feynman integral
\eq
V_T=-e^2\int\frac{d^Dl}{i(2\pi)^D}\, S_\pm(p_1+l)S_\pm(p_2-l)\,
\frac{(2p_1+l)^i(2p_2-l)^j}{l^2}\, 
\biggl(\delta^{ij}-\frac{l^il^j}{{\bf l}^2}\biggr)\, .
\en
Below, instead of $V_T$, we shall present the result for $V_C+V_T$.
Applying threshold expansion, it can be shown that the sum of
Coulomb and transverse vertices can be rewritten as
\eq
V_C+V_T=e^2\int\frac{d^Dl}{i(2\pi)^D}\, S_\pm(p_1+l)S_\pm(p_2-l)\,
\frac{4p_1p_2+l^2}{l^2}\, .
\en
Using the same technique as in the case of the Coulomb vertex, 
we finally obtain a compact expression valid to all orders in the $\e$-expansion,
\begin{align}
V_C+V_T &=
- \frac{2\alpha\, p_1p_2}{M_\pi\sqrt{s}}\biggl(
\frac{\pi M_\pi}{4q_0} + i \theta_c(q_0)\biggr)
+\frac{i\alpha q_0}{2\sqrt{s}}+\Order(d-3)\, \nnnl
&= -
\alpha \biggl\{
\frac{\pi(1+\sigma^2)}{4\sigma}+i \biggl(\frac{1+\sigma^2}{\sqrt{1-\sigma^2}} \, \theta_c(q_0)
-\frac{\sigma}{4} \biggr) \biggr\}
+\Order(d-3)\, ,\label{eq:VC+VT}
\end{align}
where in the second line, we have introduced 
$\sigma=\sqrt{1-4M_\pi^2/s}$.
As in the Coulomb vertex, the photon 3-momentum
in the loop ${\bf l}$ counts as $\Order(\epsilon)$.
Note that $V_C+V_T$ is explicitly Lorentz-covariant. 
Equation~\eqref{eq:VC+VT} can also be derived in the formalism of Refs.~\cite{Goity,Lehmann}, 
see Ref.~\cite{BKthesis}.

\subsection{Transverse vertex in crossed channel}\label{sec:xedtransvert}

The vertex in which one of the photon lines is attached to the incoming
kaon line is depicted in Fig.~\ref{fig:demo}d. Neglecting all factors
in the numerator, we consider the following scalar integral
\eq
V_K=e^2\int\frac{d^Dl}{i(2\pi)^D}\,S_\pm(p_1-l)S_K(p_K-l)\frac{1}{-l^2}\, .
\en
After integrating over $l^0$ we find
\eq\label{eq:VK}
V_K=e^2\int\frac{d^dl}{(2\pi)^d}\,\frac{1}{2w_1'2w_K'}\,\frac{1}{2|{\bf l}|}\,
\frac{1}{(w_1'-w_1+|{\bf l}|)(w_K'-w_K+|{\bf l}|)}\, ,
\en
where $w_1'=w_\pm({\bf p}_1-{\bf l})$,
$w_1=w_\pm({\bf p}_1)$, $w_K'=w_K({\bf p}_K-{\bf l})$ and
$w_K=w_K({\bf p}_K)$.
{}From Eq.~\eqref{eq:VK} one concludes that the photon momentum 
${\bf l}$ has to be counted as $\Order(\epsilon^2)$. However, since
\eq
w_1'-w_1+|{\bf l}|=|{\bf l}|-\frac{{\bf p}_1{\bf l}}{w_1}+\ldots\, ,
\quad\quad
w_K'-w_K+|{\bf l}|=|{\bf l}|-\frac{{\bf p}_K{\bf l}}{w_K}+\ldots\, ,
\en
and since the second term in this expansion counts as $\Order(\epsilon^3)$, we arrive
at no-scale integrals. Consequently, we conclude that $V_K=0$;
see also the discussion in Appendix~\ref{app:infra-ultra}.

\subsection{Coulomb exchange in strong loops}\label{sec:Coulombloop}

The exchange of a Coulomb photon  in the strong loop (see 
Fig.~\ref{fig:demo}e) is described by the following integral:
\be
J_C = e^2\int\frac{d^Dl}{i(2\pi)^D}\,\frac{d^Dk}{i(2\pi)^D}\,
S_\pm(l)S_\pm(Q-l)S_\pm(k)S_\pm(Q-k)\,
\frac{(w_1'+w_1'')(w_2'+w_2'')}{|{\bf l}-{\bf k}|^2}\, ,
\ee
where $Q=p_1+p_2$, $w_1'=w_\pm({\bf l})$, $w_2'=w_\pm({\bf Q}-{\bf l})$, 
$w_1''=w_\pm({\bf k})$, and $w_2''=w_\pm({\bf Q}-{\bf k})$. Replacing
$(w_1'+w_1'')(w_2'+w_2'')=2w_1w_2+\Order(\epsilon^2)$, where $w_1=w_\pm({\bf p}_1)$
and $w_2=w_\pm({\bf p}_2)$, integrating over 
$l^0$, $k^0$ and re-defining the integration 3-momenta ${\bf l}$, ${\bf k}$
along similar lines as in Sect.~\ref{sec:Coulombvertex},
we finally find
\eq\label{eq:Jc}
J_C=\frac{e^2 M_\pi^2}{s}\,\int\frac{d^dl}{(2\pi)^d}\,\frac{d^dk}{(2\pi)^d}\,
\frac{1}{{\bf l}^2-q_0^2}\,\frac{1}{|{\bf l}-{\bf k}|^2}\,\frac{1}{{\bf k}^2-q_0^2}+\ldots\, ,
\en
where $q_0^2=s/4-M_\pi^2$ and $s=Q^2$. From the above expression it is seen 
that ${\bf l},{\bf k}$ count as $\Order(\epsilon)$. The higher-order
corrections in Eq.~\eqref{eq:Jc} do not change the location of the threshold
given by the relativistically invariant relation $s=4M_\pi^2$.

Calculating the integral in Eq.~\eqref{eq:Jc}, we  arrive at
\eq
J_C=-\frac{\alpha M_\pi^2 }{8\pi s}\,\biggl(16\pi^2\Lambda
+ \ln \Bigl( -\frac{4q_0^2}{\mu^2}\Bigr) \biggr)+\ldots\, ,
\en
where
\eq
\Lambda = \frac{\mu^{2(D-4)}}{16\pi^2}\biggl(\frac{1}{D-4}-\Gamma'(1)-\ln 4\pi-1\biggr)\, .
\label{eq:JC1}
\en

\section{Power counting}\label{sec:powercounting}

We start by summarizing the power counting rules for $K\to 3\pi$ decays without photons.
\begin{enumerate}
\item Polynomial interaction terms can be organized in even powers of $\e$, 
starting at $\e^0$.  The inclusion of higher-order derivative terms in loop
diagrams increases the power of $\e$ of the respective graphs accordingly.
\item In loop diagrams, each pion propagator counts as $\e^{-2}$, each loop integration as $\e^5$
(as for an integration momentum $l$,  $d\mathbf{l}=\Order(\e)$ and $dl^0=\Order(\e^2)$).
\item Each loop formed with two-body rescattering terms (counted as $\Order(a)$)
therefore increases the power of $\e$ by $\e^5 (\e^{-2})^2 = \e^1$, therefore
one-loop diagrams are $\Order(a^1\e^1)$, two-loop diagrams $\Order(a^2\e^2)$, etc.
Loops with three-body terms increase the power of $\e$ by $(\e^5)^2 (\e^{-2})^3 = \e^4$.
\end{enumerate}
The examples of the preceding section show how these rules have to be amended
in the presence of (virtual) photons. 
Note that, as $\mpc^2-\mpn^2 = \Order(e^2)$ (when neglecting a tiny contribution
of second order in the light quark mass difference), one might infer that
the electric coupling should be included in a unified power counting, 
where $e=\Order(\e)$.  However, we do not follow this approach, but instead
count $e$ as a separate expansion parameter, such that we will altogether 
have a three-fold expansion in $\e$, $a$, and $e$.
In the following, we have to distinguish between
``soft'' photons ($l_0=\Order(\e^2)$, $\mathbf{l}=\Order(\e)$) and
``ultrasoft'' photons ($l_0=\Order(\e^2)$, $\mathbf{l}=\Order(\e^2)$).
\begin{enumerate}
\item[4.]
Coulomb photons are always soft, hence their propagators are
$\Order(\e^{-2})$, and loop integration still counts as $\e^5$.
Furthermore, the couplings to bosons are of $\Order(\e^0)$.
The addition of a Coulomb photon to a ``skeleton'' diagram,
consisting only of interaction vertices without electromagnetism, hence changes the 
non-relativistic power of the diagram by a factor of $\e^5 (\e^{-2})^3=\e^{-1}$. 
The two most relevant examples for $K\to 3\pi$ decays discussed individually above 
are the vertex correction with a Coulomb photon, Sect.~\ref{sec:Coulombvertex}, 
and Coulomb exchange within charged pion loops, Sect.~\ref{sec:Coulombloop}.
\item[5.] Soft transverse photons have couplings of $\Order(\e^1)$, hence
the exchange of transverse photons is suppressed by two orders with respect to
the corresponding Coulomb exchange diagram.
An example for this is given in Sect.~\ref{sec:transvert}.
\item[6.] For ultrasoft transverse photons, with all momentum components of order $\e^2$,
the propagator counts as $\e^{-4}$ and loop integration as $\e^8$.  
Therefore, adding an ultrasoft transverse photon to a skeleton diagram changes 
its power counting by a factor of (taking care of loop integration, boson propagators,
photon propagator, and photon vertices) $\e^8 (\e^{-2})^2 \e^{-4} \e^2 = \e^2$.
Examples are the pion self-energy (Sect.~\ref{sec:selfenergy}) and the exchange
of a transverse photon in the crossed channel (Sect.~\ref{sec:xedtransvert}).
\end{enumerate}
As the addition of a Coulomb photon modifies the power counting of a given
diagram by a factor of $e^2/\e$, it is obvious that we can consider \emph{very} small
momenta for which this ratio ceases to be a small parameter.  This is precisely what
happens in the energy region where pionium bound states are formed and will
be discussed in detail in Sect.~\ref{sec:pionium}.  The scheme formulated above
holds outside a small energy window around the pionium singularities, where then
the standard perturbative treatment of electromagnetism is valid.

As the inclusion of real photon radiation effects takes place on the 
level of widths or decay spectra, we briefly discuss the non-relativistic
power counting for both non-radiative and radiative widths/decay spectra.
The non-radiative differential decay width $d\Gamma$ for $K\to 3\pi$ is schematically given by
\be
d\Gamma ~\propto~ \Bigl[\prod  d\mu(p_i) \Bigr]
\delta^{(4)}\Bigl(P-\sum p_i\Bigr) |\M|^2 ~,~~
d\mu(p_i) = \frac{d^3p_i}{2p_i^0(2\pi)^3} ~.
\ee
Each invariant phase space element counts according to
$d\mu(p_i) \propto \e^3$, while the energy- and momentum-conserving
$\delta$-function induces a factor of $\e^{-5}$. 
With the matrix element $\M$ starting at $\Order(\e^0)$, it is therefore 
obvious that the non-radiative differential decay width starts out as $\Order(\e^4)$, 
and the decay spectrum with respect to $s_3$ obeys
\be
\frac{d\Gamma}{ds_3} ~=~ \Order(\e^2) ~.
\ee

The radiative differential decay width $d\Gamma_\gamma$ is schematically given by
\be
d\Gamma_\gamma ~\propto~ \Bigl[\prod  d\mu(p_i) \Bigr] d\mu(k)
\delta^{(4)}\Bigl(P-\sum p_i-k\Bigr) |\M_\gamma|^2 ~,~~
d\mu(k) = \frac{d^3k}{2k^0(2\pi)^3} ~. \label{eq:dGgamma}
\ee 
Real photons necessarily are ultrasoft,
as the energy component is $k^0 = \Order(\mk-3\mpi) = \Order(\e^2)$, and the on-shell condition 
enforces the same order for the three-momentum components.
Therefore
the additional integration over the photon phase space is $d\mu(k)=\Order(\e^4)$.
The non-relativistic power counting for the radiative matrix element 
is most easily seen in the Coulomb gauge, 
where the matrix element for photon radiation (with momentum $k$) off the external legs
(with charges $Q_i$, masses $M_i$, and momenta $p_i$)
is given by
\be
\M_\gamma ~\propto~ e \sum  Q_i\frac{\mathbf{p}_i\boldsymbol{\e}^*}{M_i k^0} 
\bigl(1+\Order(\e)\bigr)
~=~ \Order\Bigl(\frac{e}{\e}\Bigr)~.\label{eq:Mgamma}
\ee
Combining Eqs.~\eqref{eq:dGgamma}, \eqref{eq:Mgamma}, we obtain
$d\Gamma_\gamma = \Order(e^2\e^6)$ and
\be
\frac{d\Gamma_\gamma}{ds_3} ~=~ \Order(e^2\e^4) ~.
\ee
We find therefore that, for the projected accuracy, it is sufficient to 
calculate bremsstrahlung effects at \emph{leading} non-vanishing order in the 
non-relativistic expansion.  

The following conclusions can furthermore be drawn.
\begin{enumerate}
\item As the relation between the (massive boson) propagators in the relativistic
and the non-relativistic theories is given by
\[
\frac{1}{p^2-M^2} ~=~ \frac{1}{2\omega(\mathbf{p})(p^0-\omega(\mathbf{p}))} + \Order(\e^0) ~,
\]
a calculation of the real photon radiation effects in both theories is equivalent
up to the desired order.  We can therefore simply expand the relativistic results
in $\e$.
\item Contributions to the radiative width that are not of bremsstrahlung type, 
i.e.\ the photon is not attached to one of the external legs, is further suppressed
in the $\e$-expansion and does not need to be considered.
\item The same goes for bremsstrahlung type contributions with derivative vertices
or loops, which are also of subleading order in the $\e$-expansion.
\end{enumerate}

At the end of this section, we can summarize our conclusions from the above 
considerations by discussing the Feynman graphs for all real and virtual photon corrections
necessary for a consistent calculation of $d\Gamma/ds_3$
up-to-and-including $\Order(e^2 a^0 \e^4)$ (all channels) and $\Order(e^2 a^1 \e^2)$ 
(for the ``main modes'' only). 

\begin{figure}
\centering
\includegraphics[width=\linewidth]{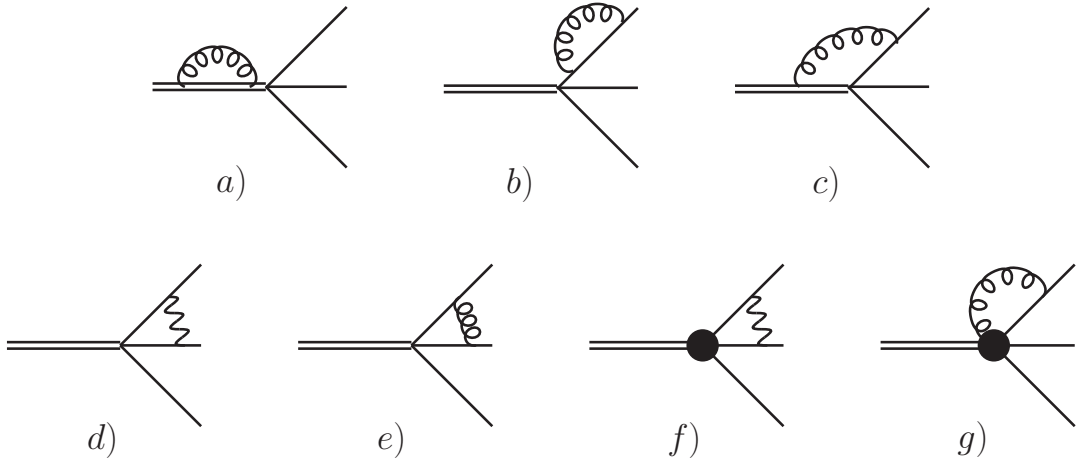}
\caption{Representative set of one-loop diagrams for $K\to 3\pi$ including a virtual photon.
Double, single, wiggly, and curly lines denote kaons, pions, Coulomb, and 
transverse photons, respectively.
The thick blob denotes an interaction vertex
with two derivatives.}\label{fig:oneloop}
\end{figure}
 Virtual photon corrections suppressed with respect to leading tree terms by 
factors up to $\Order(e^2 \e^2)$, i.e.\ without any 
 $\pi\pi$ rescattering contributions, are shown in Fig.~\ref{fig:oneloop}.
According to the counting rules spelt out above, 
 the transverse photon diagrams Fig.~\ref{fig:oneloop}a--c are all of the maximal
order we consider here, so no further derivative vertices have to be taken into
 account for the same diagram topologies.
We need to consider both diagram Fig.~\ref{fig:oneloop}d for Coulomb
 ($\Order(e^2\e^{-1})$) and Fig.~\ref{fig:oneloop}e for transverse ($\Order(e^2\e^1)$) photons.
 Coulomb exchange also necessitates the calculation of the same
 graph with a two-derivative vertex, Fig.~\ref{fig:oneloop}f, which
is then also of $\Order(e^2\e^1)$.
 We furthermore wish to point out that no diagrams with photons coupling to 
interaction vertices, see Fig.~\ref{fig:oneloop}g for an example, need to be considered,
 as those necessarily contain transverse photons with derivative interactions
and therefore only start to contribute at $\Order(e^2 \e^4)$.

\begin{figure}
\centering
\includegraphics[width=0.75\linewidth]{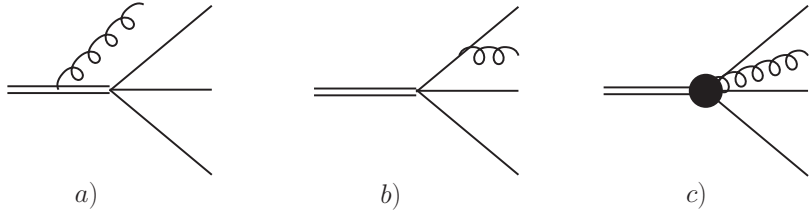}
\caption{Diagrams for the bremsstrahlung process $K\to 3\pi\gamma$.
The line style is as in Fig.~\ref{fig:oneloop}.
}\label{fig:brems}
\end{figure}
The bremsstrahlung diagrams are displayed in Fig.~\ref{fig:brems}.  
As argued above, only their leading order in $\e$ contributes to $d\Gamma/ds_3$
at $\Order(e^2 \e^4)$, therefore no diagrams with additional derivative 
vertices are needed.  In particular, structure-dependent photon radiation
as in the graph Fig.~\ref{fig:brems}c is suppressed by the absence of a 
meson propagator and the derivative in the vertex by at least two orders in $\e$.

\begin{figure}
\centering
\includegraphics[width=\linewidth]{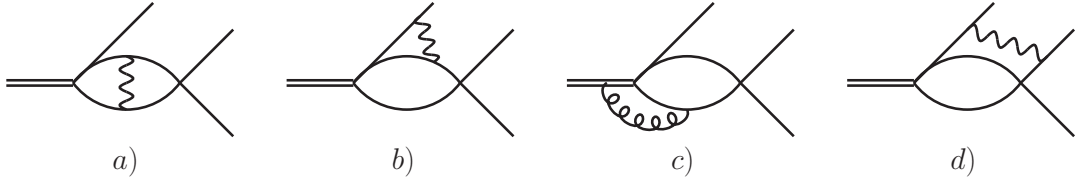}
\caption{Two-loop diagrams for $K\to 3\pi$, including a virtual photon
and one $\pi\pi$ rescattering vertex.
Line style as in Fig.~\ref{fig:oneloop}.
}\label{fig:twoloop}
\end{figure}
Finally, we turn to the diagrams that contain one further $\pi\pi$ interaction vertex, 
i.e.\ that are of $\Order(e^2a^1\e^0)$.  These are displayed in Fig.~\ref{fig:twoloop}.
Note that we only consider these corrections for the two ``main modes'' 
$K_L \to 3\pi^0$ and $K^+ \to \pi^0 \pi^0\pi^+$.  
Obviously, the diagram Fig.~\ref{fig:twoloop}a 
(discussed extensively in Sect.~\ref{sec:Coulombloop} and below in Sect.~\ref{sec:pionium}) 
is the only one contributing to the former, 
and neither of the ``main modes'' features the topology shown in Fig.~\ref{fig:twoloop}d
(which requires at least two charged pions in the final state).  
As the pion loop as such is of $\Order(\e)$, only the addition of a Coulomb photon
can lead to the required order in the non-relativistic counting, 
and transverse photon contributions like Fig.~\ref{fig:twoloop}c are of higher order
and can be neglected.
Finally, the graph shown in Fig.~\ref{fig:twoloop}b could potentially contribute
to $K^+ \to \pi^0 \pi^0\pi^+$, with a $\pi^+\pi^-$ pair in the loop rescattering into 
$\pi^0\pi^0$.  However, it is easily seen that the two contributions with the 
photon hooked to the $\pi^+$ and the $\pi^-$ in the loop exactly cancel, hence we 
do not have to consider this diagram either.
We conclude that the graph discussed in Sect.~\ref{sec:Coulombloop} is indeed the only one
at $\Order(e^2a^1\e^0)$ that needs to be taken into account.

\section{Pionium}\label{sec:pionium}

In previous sections, we have analyzed the effect of radiative corrections in the
 perturbative regime, where an expansion in the fine structure constant $\alpha$ makes sense. 
In that region, pion momenta are counted as order $\epsilon$. Here, 
we investigate a region in momentum space where this is not necessarily true anymore: very near 
the cusp, pionium can form, and an expansion in $\alpha$ is obviously not possible. 
The three-momenta of pions 
are of  order    $\alpha$, and one needs to take into account an infinite number of graphs 
to describe this momentum regime properly. 

There are two possibilities to cope with this problem. Either, one attempts to set up a formalism
 that allows one to treat the whole decay region, including the formation of pionium, in a coherent 
and consistent manner. Or, one proceeds in the manner 
in which the data analysis was performed by the NA48/2
 collaboration~\cite{Batley}: one excludes a region around the cusp where this phenomenon occurs, 
and restricts the analysis  to the region where a perturbative expansion in $\alpha$ is
 possible. 
We adopt this route in this article, for the following reason. Even if
 one might be able to set up a coherent formalism for the whole decay region,\footnote{Note that, 
 at present, there is no consent with respect to
 the pionium contribution 
 in $K\to 3\pi$ decays. For example, it was shown that, including this
 contribution, a better fit
 to the data  can be achieved in the vicinity of the cusp~\cite{Batley}.
 However, the best fit value of the
 $K^\pm\to \pi^\pm+\mbox{pionium}$ decay rate normalized
 to the $K^\pm\to \pi^\pm\pi^+\pi^-$ decay rate
 is equal to $(1.61\pm 0.66)\times 10^{-5}$ and is
 considerably higher than the theoretical prediction $\sim 0.8\times
 10^{-5}$~\cite{Silagadze,Batley}.
 Furthermore,
it has recently been argued~\cite{Wycech1,Wycech2} that this value
 substantially increases if the interference effect with the
$\pi^0\pi^0$ intermediate states is taken into account,
but the correction is model-dependent. A complete analysis of the
problem is still pending~\cite{Wycech1,Wycech2}.}
there would be no  real benefit of such a tour de force in the present context: as we are interested in a 
precise extraction of the $\pi\pi$ scattering lengths, it is sufficient to stay in a region in 
momentum space where
a perturbative treatment is possible, and where the effect of the cusp still allows one to extract the 
scattering lengths with high precision. Investigating the 
region where pionium is formed would then not provide any new information on the scattering
 lengths, but merely test the formalism to describe that region in a proper manner. This is so, because 
the energy resolution in 
present investigations of $K\to 3\pi$ decays does by far not suffice to e.g.\ measure the width 
of the ground state of pionium, which is an alternative possibility to measure $\pi\pi$ 
scattering lengths.  In our opinion, this alternative way 
 is best covered by dedicated
 measurements of the production and decay of pionium as performed e.g.\ in the DIRAC 
experiment~\cite{dirac1,dirac2}. Here,
 the theoretical background for the data analysis has already been provided with the necessary 
precision~\cite{nrqft6,nrqft8,nrqft9,nrqft10}.

In the data analysis of NA48/2~\cite{Batley},
a region around the cusp was omitted,
\bea\label{eq:cutNA48}
s_i\not\in[4\mpc^2-\Delta_s,4\mpc^2+\Delta_s ]\,\,;\,\,\quad \Delta_s = 
5.25\times 10^{-4} \GeV^2\,\,,
\eea
whenever $s_i$ corresponds to a channel where pionium can be formed.
It therefore remains to answer the following question: is the region outside this interval
accessible to a perturbative expansion in $\alpha$? As Ref.~\cite{Batley} has already demonstrated that
this region does allow a precise measurement of the $\pi\pi$ scattering length 
(in the absence of radiative corrections), one is done 
once one knows that this question can be answered in a positive sense. The present section 
is devoted to an investigation of this problem.

\subsection[$K_L\to 3\pi^0$]{\boldmath{$K_L\to 3\pi^0$}}

\begin{figure}
\centering
\includegraphics[width=0.5\linewidth]{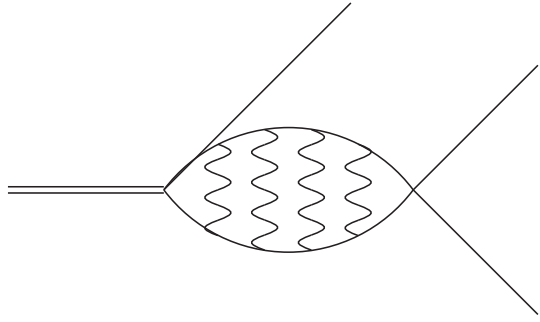}
\caption{Diagram with  four Coulomb photons exchanged between the oppositely charged pions. 
The quantity $J_{nC}$ in Eq.~\eqref{eq:schwinger} describes the sum of diagrams with
 $n\geq 2$ Coulomb photon 
exchanges.}
\label{fig:schwinger}
\end{figure}

It is useful to start with the decay $K_L\to 3\pi^0$, because,
as was shown at the end of the previous section,
the graph Fig.~\ref{fig:twoloop}a is the only one that needs to be taken 
into account in the perturbative region at order $\Order(e^2a^1\epsilon^0)$. It was 
evaluated in the present framework 
in Sect.~\ref{sec:Coulombloop}.
In the non-perturbative region, at order $a$, there are additional 
Coulomb exchanges  that must be taken into account, see Fig.~\ref{fig:schwinger} for an example.
The effect of these Coulomb exchanges is to modify the one-loop 
integrals $J_{+-}(s)$ displayed in Eq.~\eqref{eq:Jab}, in the following 
manner,
\bea\label{eq:replace}
 J_{+-}(s)&\to&J_{+-}(s)+J_C(s)+J_{nC}(s)\,.
\eea
Here, the second term on the right-hand side denotes the  contribution of Fig.~\ref{fig:twoloop}a. 
At the order considered here, it is given by
\bea
J_C(s)=-\frac{\alpha}{32\pi}\left(16\pi^2\Lambda+\ln{\Bigl(-\frac{4q_0^2}{\mu^2}\Bigr)}\right)\,
;\, q_0^2=\frac{s}{4}-M_\pi^2\,.
\eea
The third term on the right-hand side of Eq.~\eqref{eq:replace} stands for the sum of 
$n\geq 2$ Coulomb photon exchanges~\cite{schwinger},
\bea\label{eq:schwinger}
J_{nC}&=&-\frac{\alpha}{16\pi}\left(\Psi(1-\nu)-\Psi(1)\right)\,,\nonumber\\
\nu&=&\frac{\alpha M_\pi}{2\sqrt{-q_0^2}}\,,\,
\Psi(x)=\frac{d}{dx}\ln{\Gamma(x)}\,.
\eea
Two remarks are in order. First, the ultraviolet contribution that shows up in
$J_C(s)$ can be absorbed in the  couplings at tree level. In the following, we set the scale at $\mu=M_\pi$, and use instead of $J_C$ the expression
\bea\label{eq:replaceJc}
J_C\to\bar J_C= -\frac{\alpha}{32\pi}\ln{\left(-\frac{4q_0^2}{M_\pi^2}\right)}\,.
\eea
The choice of the scale is irrelevant: a change of scale simply induces 
a corresponding change in the tree-level couplings. Second, we performed the evaluation 
of $J_{nC}$ in the non-relativistic framework described in Ref.~\cite{nrqft7,physrep}, which differs from 
the one proposed here by use of propagators where the non-relativistic expansion
$w_\pm(\mathbf{p}) \to M_\pi + {\mathbf{p}^2}/{2M_\pi}+\cdots$ is performed.
A calculation with the propagators used here would be rather demanding. On the other hand, 
that calculation would lead to the same result at 
leading order in the momentum expansion. As we shall drop the contributions from 
multi-Coulomb exchanges anyhow, see below, we do not attempt to perform this considerably 
 more complicated calculation.

The effects of the 
non-perturbative regime is clearly seen in the multi-Coulomb contribution 
$J_{nC}$. Indeed, once the pion momenta count as order $\alpha$, $\nu$ is of order 1, and a 
perturbative expansion is no longer possible. A result of this  are the pionium poles 
that show up in $J_{nC}$,  at
\bea
q_0^2=-\frac{\alpha^2M_\pi^2}{4n^2}\,\,\,;\,\,n=1,2,\ldots\,.
\eea
On the other hand, this phenomenon only happens for pion momenta that are sufficiently 
small. Expanding $\Psi(1-\nu)$ around $\nu=0$, it is seen that multi-Coulomb exchanges are 
certainly negligible in comparison to the leading term $\bar J_C$  in case that the width 
of the cut satisfies $\Delta_s\geq 10^{-4}$ GeV$^2$. As a result of this,
Coulomb exchanges finally amount to the replacement
\bea\label{eq:replacementCoulomb}
 J_{+-}(s)&\to& \JC(s) = J_{+-}(s)+\bar J_C(s)\,.
\eea
In the following, only the combination $\JC(s)$ will appear.

\begin{figure}
\centering
\includegraphics[width=0.8\linewidth]{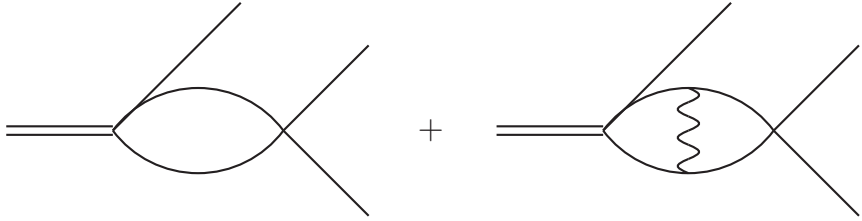}
\caption{A contribution to the ratio $R$ in Eq.~\eqref{eq:ratioR},
evaluated with the replacement Eq.~\eqref{eq:replacementCoulomb} at one-loop order.
The pertinent  $R$ is displayed in Fig.~\ref{fig:ratioR}
with a solid line.}
\label{fig:graphsRa}
\end{figure}

\begin{figure}
\centering
\includegraphics[width=0.8\linewidth]{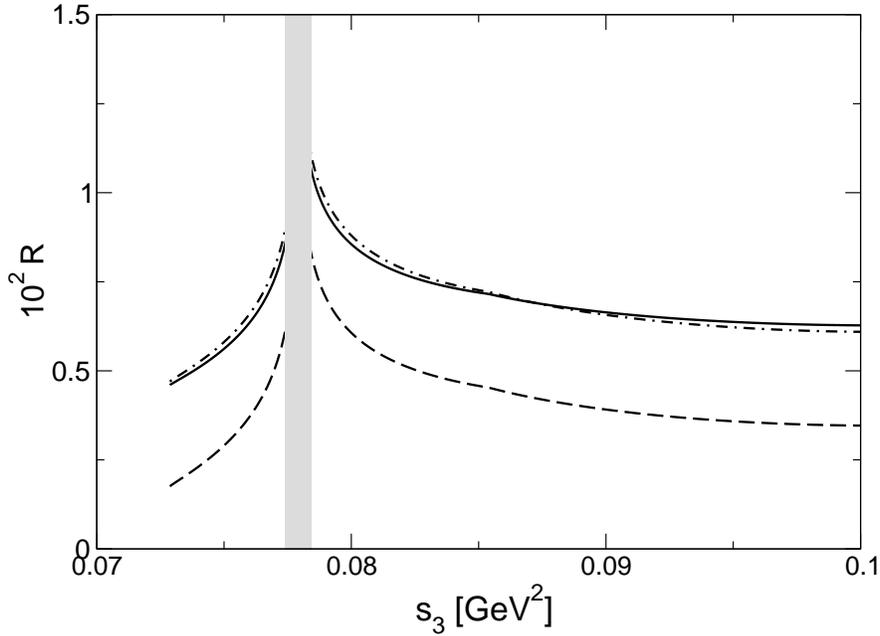}
\caption{The effect of one-Coulomb exchanges in the spectrum of
 $K_L\to 3\pi^0$ decays.
Shown is the ratio $R$ from Eq.~\eqref{eq:ratioR}. 
The grey band denotes the region Eq.~\eqref{eq:cutNA48}, 
excluded in the data analysis of 
Ref.~\cite{Batley}. The solid line is evaluated 
with the replacement Eq.~\eqref{eq:replacementCoulomb} at one-loop order.
The dashed line corresponds to the same ratio, 
with the replacement Eq.~\eqref{eq:replacementCoulomb} in bubble graphs
 at order $a$ and $a^2$. 
The dashed-dotted line illustrates that the effect of Coulomb exchanges 
at order $\alpha a^2$ can be  removed to a large extent 
through a renormalization of the coupling $K_0$ in Eq.~\eqref{eq:lag_KL}. 
 }
\label{fig:ratioR}
\end{figure}

We now investigate the size of the one-Coulomb 
exchange at order $a$ and $a^2$.  
We drop contributions from derivative interactions in the matrix element
 and evaluate the ratio
\bea\label{eq:ratioR}
R=\frac{
\frac{d\Gamma}{ds_3}^{\hspace{-.8mm}\rm{int}} - \frac{d\Gamma}{ds_3}^{\hspace{-.8mm}0}
               }
{
\frac{d\Gamma}{ds_3}^{\hspace{-.8mm}0}
}\,.
\eea
The quantity $\frac{d{\Gamma}}{ds_3}^{\hspace{-.8mm}\rm{int}}$ is calculated with the 
replacement Eq.~\eqref{eq:replacementCoulomb}. We first consider the case where this replacement 
is  performed at one-loop order only.
Charged pion loops  then generate the two graphs displayed in Fig.~\ref{fig:graphsRa}. 
On the other hand, the Coulomb contributions $\bar J_C$ are dropped in 
$\frac{d\Gamma}{ds_3}^{\hspace{-.8mm}0}$. The decay spectrum is calculated
by use of the procedure described in Appendix~\ref{app:dalitzkl3pi0}.\footnote{We note that, 
by using this procedure, the number of decay events in the vicinity of the cusp is increased 
by a factor three with respect to the standard evaluation of $d\Gamma/ds_3$.} The result for the ratio 
$R$  is displayed in Fig.~\ref{fig:ratioR}, 
with a solid line, with tree-level 
couplings $L_0=-1.4 K_0$, see Eq.~\eqref{eq:lag_KL}.  The grey band denotes the region discarded
in the NA48/2 analysis. It is seen that outside this region, the correction due to 
one-Coulomb exchanges is of the order of  one percent or less.  Note that the ratio $R$ is linear
in the ratio ${L_0}/{K_0}$ to very high accuracy.


\begin{figure}
\centering
\includegraphics[width=0.8\linewidth]{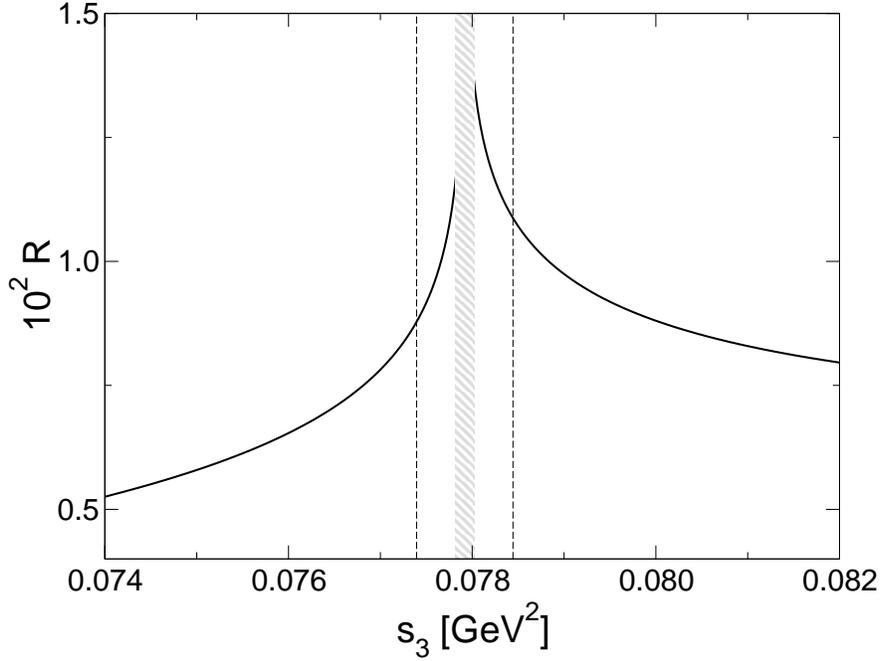}
\caption{The effect of one-Coulomb exchanges inside the region Eq.~\eqref{eq:cutNA48} cut out in the 
NA48 analysis. The solid line again shows the ratio $R$ at one-loop order, discussed in the text.
 The dashed vertical lines indicate the region Eq.~\eqref{eq:cutNA48}, whereas the hatched 
grey band denotes the same region, decreased by a factor of~5.}
\label{fig:Rclose}
\end{figure}

\begin{figure}
\centering
\includegraphics[width=0.5\linewidth]{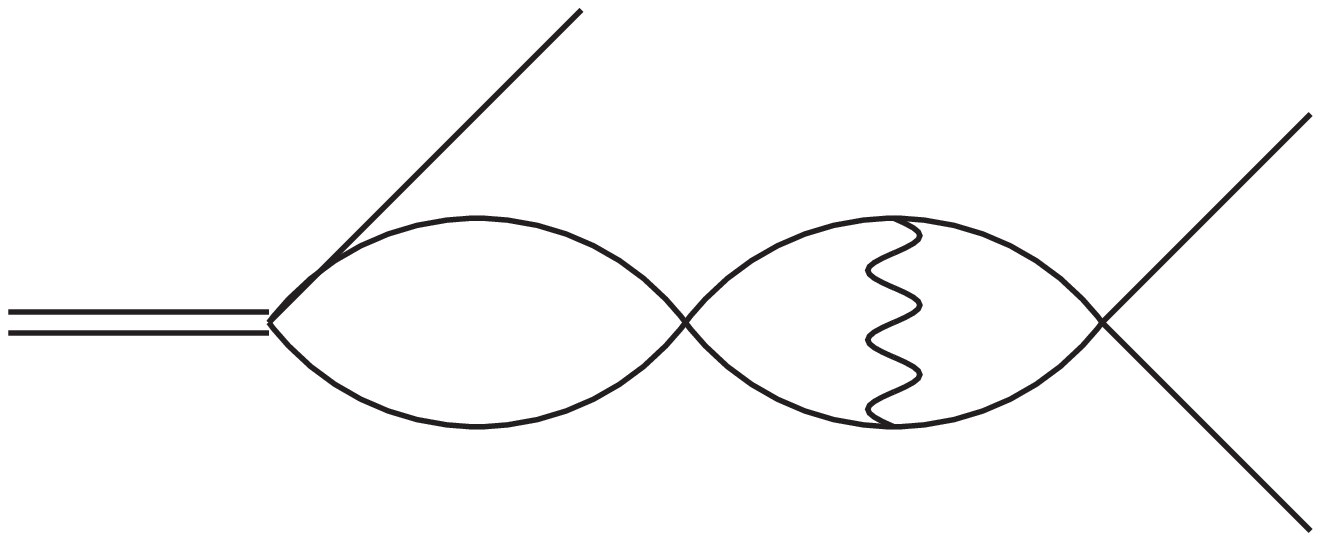}
\caption{Contributions to the ratio $R$ at two-loop order. 
In graphs with charged pion loops,  one additional Coulomb exchange is retained.
The corresponding ratio $R$ in Eq.~\eqref{eq:ratioR} is displayed
 in Fig.~\ref{fig:ratioR}, with a dashed line.}
\label{fig:graphsRa2}
\end{figure}

We may also investigate $R$ closer to the position of the pionium poles. 
In Fig.~\ref{fig:Rclose} we display
the ratio $R$ again, now also inside the region cut in the NA48/2 analysis.
The grey vertical band corresponds to $s_3\not\in [4M_\pi^2-{\Delta_s/5},
4M_\pi^2+{\Delta_s/5}]$, i.e., 
the excluded region Eq.~\eqref{eq:cutNA48} is reduced by a factor 5. 
The dashed vertical lines correspond to the original cut 
 Eq.~\eqref{eq:cutNA48}. It is seen that also in this enlarged region of
 phase space, 
the electromagnetic corrections due to one-Coulomb exchanges remain small. 

We now comment on effects at order $\Order(e^2a^2\epsilon)$ -- a contribution at this order 
is e.g.\ generated by the graph displayed in Fig.~\ref{fig:graphsRa2}.
A full calculation  of all pertinent graphs would be very complicated. For
example, one would have to consider Coulomb photon exchanges 
in the two-loop graph displayed in 
Fig.~\ref{fig:2loop_typ}a  -- a tremendous 
calculation. In order 
to have at least an order-of-magnitude  estimate of the effect of these
contributions, 
we  perform the replacement Eq.~\eqref{eq:replacementCoulomb} in  bubble graphs  at one- and two-loop order. 
The result for the ratio $R$ is shown in Fig.~\ref{fig:ratioR}, with a dashed line.
It is seen that the change in $R$  is rather substantial. However, almost all of  this change can be 
absorbed in a redefinition of the coupling $K_0$ in Eq.~\eqref{eq:lag_KL}. 
This is illustrated with the dash-dotted line in the same figure, which results 
from the dashed one by changing   $K_0$ by half a percent, and which illustrates that the apparent 
sensitivity on the terms of order $\alpha a^2$ is of no concern in this region of momentum space.

\subsection[$K^+\to\pi^0\pi^0\pi^+$]{\boldmath{$K^+\to\pi^0\pi^0\pi^+$}}
The discussion in this channel is analogous to the purely neutral channel just considered. 
We have checked that the corresponding ratio $R$ -- see Eq.~\eqref{eq:ratioR} -- has qualitatively the same
 behavior as is shown in Figs.~\ref{fig:ratioR} for the purely neutral channel, although 
with smaller absolute value.  Furthermore, a renormalization of the tree-level coupling again
allows one to remove most of the  contributions at order $\alpha a^2$.

We come to the following conclusions.
The Coulomb corrections in hadronic loops are perturbative 
{\em at least} in the region which is used in the present data analysis by NA48/2 
collaboration (7 energy bins discarded around the threshold). One may safely
neglect  pionium formation here. Moreover, taking into account
the $\Order(e^2a\epsilon^0)$ correction suffices: the two-loop correction can
be effectively removed by a finite renormalization of 
$K\to 3\pi$ tree-level couplings. For completeness, we nevertheless display
in the following section the amplitudes with Coulomb corrections
 partially included also at two loops. According to the above discussion, 
we expect that the inclusion of these additional terms 
 does not introduce a significant modification of the extracted
 values of the scattering lengths.

\section{Results: amplitudes and correction factors}\label{sec:results}

In this section, we collect the main results of our investigation:
the correction factors $\Omega$ and the modified matrix elements
$\M^{\rm int}$ for all four $K \to 3\pi$ channels.
 
\subsection[$K_L\to 3\pi^0$]{\boldmath{$K_L\to 3\pi^0$}}

We start with the simplest case, with no external charged particles. 
Consequently, in the decomposition
\be
\frac{d\Gamma}{ds_3} \biggr|_{E_\gamma<\Emaxn} = ~
\Omega_{000} \, \frac{d\Gamma^{\rm int}}{ds_3} ~,
\ee
where $\Emaxn$ is the maximum photon energy in this channel,
the external correction factor is just equal to one, $$\Omega_{000}=1 ~.$$
$d\Gamma^{\rm int}$ is given as 
\be
d\Gamma^{\rm int} = N \, d \Phi_3 |\M_{000}^{\rm int}|^2 
\ee
(see Appendix~\ref{app:notation} for the Lorentz-invariant phase space
element $N\,d\Phi_3$), where the matrix element $\M_{000}^{\rm int}$ can be obtained
from the matrix element $\M_{000}$ in Ref.~\cite{KlongLetter} by the 
replacement
\be
J_{+-}(s_i) \to \JC(s_i) ~,
\ee
see Eq.~\eqref{eq:replacementCoulomb},
which leads to the explicit form at two-loop order
\newcommand{\xone}{(p_1^0-\mpn)}
\newcommand{\xtwo}{(p_2^0-\mpn)}
\newcommand{\xthree}{(p_3^0-\mpn)}
\begin{align}
\M_{000}^{\rm int} &= \M_0^{\rm tree}
                 +\M_0^{\rm 1-loop}
                 +\M_0^{\rm 2-loop} ~, \nonumber\\
\M_0^{\rm tree} &= K_0+K_1\left[\xone^2+\xtwo^2+\xthree^2\right] ~, \nonumber\\
\M_0^{\rm 1-loop} &= B_{0}^{(1)}(s_1)J_{00}(s_1) + B_{0}^{(2)}(s_1) \JC(s_1)
+ (s_1 \leftrightarrow s_2 )+ (s_1 \leftrightarrow s_3) ~, \nonumber \\
\M_0^{\rm 2-loop} &= \M^A_0(s_1)+\M^B_0(s_1) 
+(s_1 \leftrightarrow s_2 ) + (s_1 \leftrightarrow s_3  ) ~, \nonumber \\
\M^B_0(s_1) &= C_{00}^2\, K_0\, J^2_{00}(s_1) +  4\,C_x\,C_{+-}\, L_0\, (\JC(s_1))^2 \nn
& \quad +\big(2\, C_x^2\, K_0 + 2\, C_{00}\, C_x\, L_0\big)\JC(s_1) J_{00}(s_1),
\end{align}
and the polynomials $B_{0}^{(1)}$, $B_{0}^{(2)}$ in $\M_0^{\rm 1-loop}$ as well as the genuine
two-loop contributions $\M^A_0(s_1)$ are as given in Ref.~\cite{KlongLetter}.
Note that the replacement to $\JC(s_i)$ inside the two-loop contributions
$\M^B_0(s_1)$ is in principle beyond the order of accuracy we are considering
in this work, as it introduces a correction of $\Order(e^2a^2)$, but it still represents
a valid partial higher-order contribution.

\subsection[$K_L\to \pi^+\pi^-\pi^0$]{\boldmath{$K_L\to \pi^+\pi^-\pi^0$}}\label{sec:knc}

In the decay channel $K_L\to \pi^+\pi^-\pi^0$, all radiative corrections
are contained in the correction factor
$\Omega_{+-0}$ in the decomposition
\be
\frac{d\Gamma}{ds_3} \biggr|_{E_\gamma<\Emaxn} = 
\Omega_{+-0}(s_3,\Emaxn) \frac{d\Gamma^{\rm int}}{ds_3} ~ \label{eq:Omega+-0}
\ee
(where the expression for the maximum photon energy $\Emaxn$ 
in this channel is as given in Eq.~\eqref{eq:Emax}),
as $\M^{\rm int}$ determining $d\Gamma^{\rm int}$ only contains 
electromagnetic contributions of order $e^2 a^1$, which we ignore in this channel.
$\M^{\rm int}$ is therefore 
\be \M^{\rm int}_{+-0} = \M_{+-0} 
\ee
as given directly in Ref.~\cite{KlongLetter}.
$\Omega_{+-0}(s_3,\Emaxn)$ comprises effects of virtual photon exchange
and real photon radiation, which we may write as
\be
\Omega_{+-0}(s_3,\Emaxn) = 1+\frac{\alpha}{\pi} 
\Bigl( \omega_{+-0}^{\rm virt}(s_3) + \omega_{+-0}^{\rm real}(s_3,\Emaxn) \Bigr) + \Order(\alpha^2) ~.
\label{eq:Omega+-0decomp}
\ee
The virtual photon corrections also contain ultraviolet divergences, which
are absorbed in a redefinition of the $K \to 3\pi$ coupling constants 
as defined in Eqs.~\eqref{eq:lag_K}, \eqref{eq:lag_KL}.
In principle, these are of a form
\be
L_i = \bar L_i + e^2 \tilde L_i + \Order(e^4) ~, \quad 
\tilde L_i = \tilde L_i^r(\mu) + \beta_i \lambda ~,  
\ee
(and similarly for the $G_i$, $H_i$), 
i.e.\ the couplings $L_i$ ($G_i$, $H_i$) contain ultraviolet divergences 
and a scale-dependent part proportional to $e^2$.  
We do not indicate these divergences explicitly, but remove all poles 
at $D = 4$ automatically.

\begin{sloppypar}
The Feynman graphs for the virtual-photon corrections
are displayed in Fig.~\ref{fig:oneloop}b, d--f.
In the non-relativistic framework, we find the following result,
using Eq.~\eqref{eq:VC+VT}:
\be
\mathrm{Re}\M_{+-0} = \bigl\{L_0 + L_1 (p_3^0-\mpn ) \bigr\}
\biggl[ 1 +e^2\, \frac{1+\sigma^2}{16\sigma} \biggr] ~, \label{eq:K0corr} 
\ee
with $\sigma = \sqrt{1-4\mpc^2/s_3}$.
It is sufficient to determine the real part $\mathrm{Re}\M_{+-0}$, 
as the imaginary part would only contribute in $|\M_{+-0}|^2$ by interference
with the imaginary parts of $\pi\pi$ rescattering graphs 
and hence be of order $e^2a^1$, which is beyond the desired accuracy for this channel.
Equation~\eqref{eq:K0corr} translates into
\be
\omega_{+-0}^{\rm virt}(s_3) = \frac{\pi^2(1+\sigma^2)}{2\sigma} ~.\label{eq:omega+-0virt}
\ee
\end{sloppypar}

Bremsstrahlung corrections in 
$K_L(P) \to \pi^+(p_1)\pi^-(p_2)\pi^0(p_3)\gamma(k)$
are due to the diagram Fig.~\ref{fig:brems}.
The matrix element squared reads 
\be
\abs{\M_{+-0}^\gamma }^2 = - e^2 L_0^2 \left[ \frac{\mpc^2}{(p_1
    k)^2}+\frac{\mpc^2}{(p_2 k)^2}-\frac{s_3-2\mpc^2}{(p_1 k)(p_2 k)} \right] ~,
\ee
and we have shown in Sect.~\ref{sec:powercounting} that we can perform the 
calculation of the differential decay width in its relativistic form.
It can be written as
\bea
d\Gamma &=& N\,d\Phi_4 \abs{\M_{+-0}^\gamma }^2  ~,
\eea
where the precise form of the four-particle phase space element $N\, d\Phi_4$ 
is given in Appendix~\ref{app:notation}.
The calculation of $\omega_{+-0}^{\rm real}(s_3,\Emaxn)$ is discussed
in detail in Appendix~\ref{app:brems+-0}, leading to 
\begin{align}
\omega_{+-0}^{\rm real}(s_3,\Emaxn) &= 
\frac{8}{3}\sigma^2 \biggl[ \ln\frac{2\Emaxn}{\mpc} -\frac{7}{3} + 2\sqrt{1-\delta_0} 
-2\ln\frac{1+\sqrt{1-\delta_0}}{2} \biggr] \! + \Order(\e^4,\alpha) , \nonumber\\
\delta_0 &= \frac{8\mkl(\mkl-\mpn)\mpn\Emaxn}{\lambda_0} ~. \label{eq:omega0}
\end{align}
Performing the soft-photon approximation in the phase space integration
amounts to setting $\delta_0 \to 0$ in Eq.~\eqref{eq:omega0},
see Appendix~\ref{app:softphotonanalytic}.

\subsection[$K^+\to \pi^0\pi^0\pi^+$]{\boldmath{$K^+\to \pi^0\pi^0\pi^+$}}\label{sec:kcn}

For this channel, the decomposition of the radiative corrections effects is given by
\be
\frac{d\Gamma}{ds_3} \biggr|_{E_\gamma<\Emaxp} = 
\Omega_{00+}(s_3,\Emaxp) \frac{d\Gamma^{\rm int}}{ds_3} ~.\label{eq:defOmega00+}
\ee
We start by discussing the correction factor
$\Omega_{00+}(s_3,\Emaxp)$, where $\Emaxp$ is given in Eq.~\eqref{eq:Emax}.
We write
\be
\Omega_{00+}(s_3,\Emaxp) = 1+\frac{\alpha}{\pi} 
\Bigl( \omega_{00+}^{\rm virt}(s_3) + \omega_{00+}^{\rm real}(s_3,\Emaxp) \Bigr) + \Order(\alpha^2) ~,
\label{eq:Omega00+}
\ee
in analogy to what was done for $\Omega_{+-0}$. 

\begin{sloppypar}
The Feynman graphs for the virtual-photon corrections
are displayed in Fig.~\ref{fig:oneloop}a--c.
In the non-relativistic framework however, as discussed in Sects.~\ref{sec:selfenergy}, 
\ref{sec:xedtransvert} and Appendix~\ref{app:infra-ultra}, 
all these virtual photon contributions vanish in dimensional regularization
after applying threshold expansion. Hence we find
\be
\omega_{00+}^{\rm virt}(s_3) = 0 ~,
\ee
and the correction factor $\Omega_{00+}(s_3,\Emaxp)$ is given exclusively
in terms of real photon radiation effects.
\end{sloppypar}

The radiative decay $K^+(P) \to \pi^0(p_1) \pi^0(p_2) \pi^+(p_3) \gamma(k)$
is given in terms of the squared matrix element 
\be
\abs{\M_{00+}^\gamma}^2 =-e^2G_0^2\left[\frac{\mk^2}{(Pk)^2}+\frac{\mpc^2}{(p_3k)^2}-\frac{2Pp_3}{(p_3k)(Pk)}\right]. 
\ee
Following Appendix~\ref{app:brems00+} for the calculation of
the radiative decay spectrum, one obtains
\begin{align}
\omega_{00+}^{\rm real}(s_3,\Emaxp)  &= \frac{\lambda_c}{6\mk^2\mpc^2} \biggl\{
\ln\frac{2\Emaxp}{\mpc} - 3 + \frac{2}{3}(4-\delta_c)\sqrt{1-\delta_c} 
\nonumber\\ & \qquad 
-2 \ln\frac{1+\sqrt{1-\delta_c}}{2}  \biggr\}
+\Order(\e^4,\alpha) ~, \nonumber\\
\delta_c &= \frac{8\mk(\mk-\mpc)\mpc\Emaxp}{\lambda_c} ~. \label{eq:Omegac}
\end{align}
Performing the soft-photon approximation in the phase space integration
amounts to setting $\delta_c \to 0$ in Eq.~\eqref{eq:Omegac}.

The ``internal'' photon corrections in $\M_{00+}^{\rm int}$ are again comprised
in the replacement
\be
J_{+-}(s_i) \to \JC(s_i) 
\ee
(compare Eq.~\eqref{eq:replacementCoulomb})
in the amplitude $\M_{00+}$ displayed in Refs.~\cite{CGKR,GKR,privComm},
which leads to the following explicit form at two-loop order,
\begin{align}
\M_{00+}^{\rm int} &= \M_N^{\rm tree}
                 +\M_N^{\rm 1-loop}
                 +\M_N^{\rm 2-loop}  ~, \nonumber\\
\M_N^{\rm tree} &= G_0+G_1(p_3^0-M_\pi)+G_2 (p_3^0-M_\pi)^2+
G_3(p_1^0-p_2^0)^2 ~, \nonumber \\
\M_N^{\rm 1-loop} &= B_{N1}(s_3)\JC(s_3) + B_{N2}(s_3)J_{00}(s_3)
+ \bigl\{ B_{N3}(s_1)J_{+0}(s_1) + (s_1\leftrightarrow s_2) \bigr\} ~, \nonumber \\
\M_N^{\rm 2-loop} &= \M^A_N(s_1,s_2,s_3)+\M^B_N(s_1,s_2,s_3) ~, \nonumber \\
\M^B_N &=  4H_0C_xC_{+-} \bigl(\JC(s_3)\bigr)^2
+ 2\bigl[G_0C_x^2+H_0C_xC_{00}\bigr]\JC(s_3)J_{00}(s_3) \nn
& \quad +G_0C_{00}^2J^2_{00}(s_3) 
+ \bigl\{4G_0C_{+0}^2J^{\,\,2}_{+0}(s_1)
+(s_1\leftrightarrow s_2)\bigr\} ~,
\end{align}
and the polynomials $B_{N1}$, $B_{N2}$, $B_{N3}$ in $\M_N^{\rm 1-loop}$ as well as the genuine
two-loop contributions $\M^A_N$ are as given in Refs.~\cite{CGKR,GKR,privComm}.

We conclude this subsection with the following comment. The correction factor $\Omega_{00+}$ turns out 
to be close to unity, and $d\Gamma^{\rm{int}}/ds_3\simeq d\Gamma/ds_3$. To illustrate, we define
\bea\label{eq:Rtot}
R^{\rm{tot}}=\frac{
\Omega_{00+}\frac{d\Gamma^{\rm{int}}}{ds_3}-\frac{d\Gamma}{ds_3}}{\frac{d\Gamma}{ds_3}}\fs
\eea
Again, let us perform the replacement Eq.~\eqref{eq:replacementCoulomb} 
 only in diagrams with no derivative couplings, and at one-loop order.
\begin{figure}
\begin{center}
\includegraphics[width=10cm]{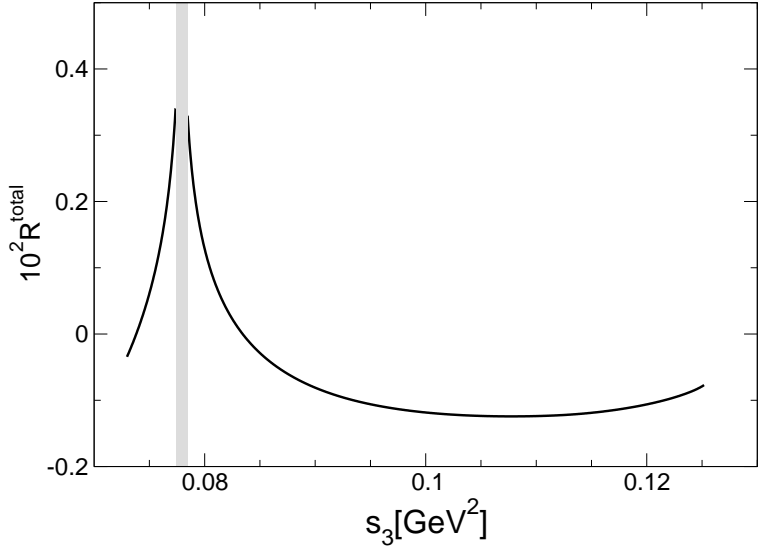} 
\end{center}
\caption{The ratio $R^{\rm{tot}}$ defined in Eq.~\eqref{eq:Rtot}, see text for more details.
 The grey band denotes the region Eq.~\eqref{eq:cutNA48}, 
excluded in the data analysis of 
Ref.~\cite{Batley}. 
It is seen that the radiative corrections are very small in this case.}
\label{fig:Rtot}
\end{figure}
We display the resulting  $R^{\rm{tot}}$ in Fig.~\ref{fig:Rtot} 
in the region defined in Eq.~\eqref{eq:cutNA48}.
We conclude from this figure that the radiative corrections to the amplitude
are small in this channel.

\subsection{On the soft-photon approximation}\label{sec:softphoton}

We wish to briefly discuss the quality of the soft-photon approximation of the
bremsstrahlung calculation in the corrections factors $\Omega_{+-0}$ and $\Omega_{00+}$
as discussed in the previous two subsections.
\begin{figure}
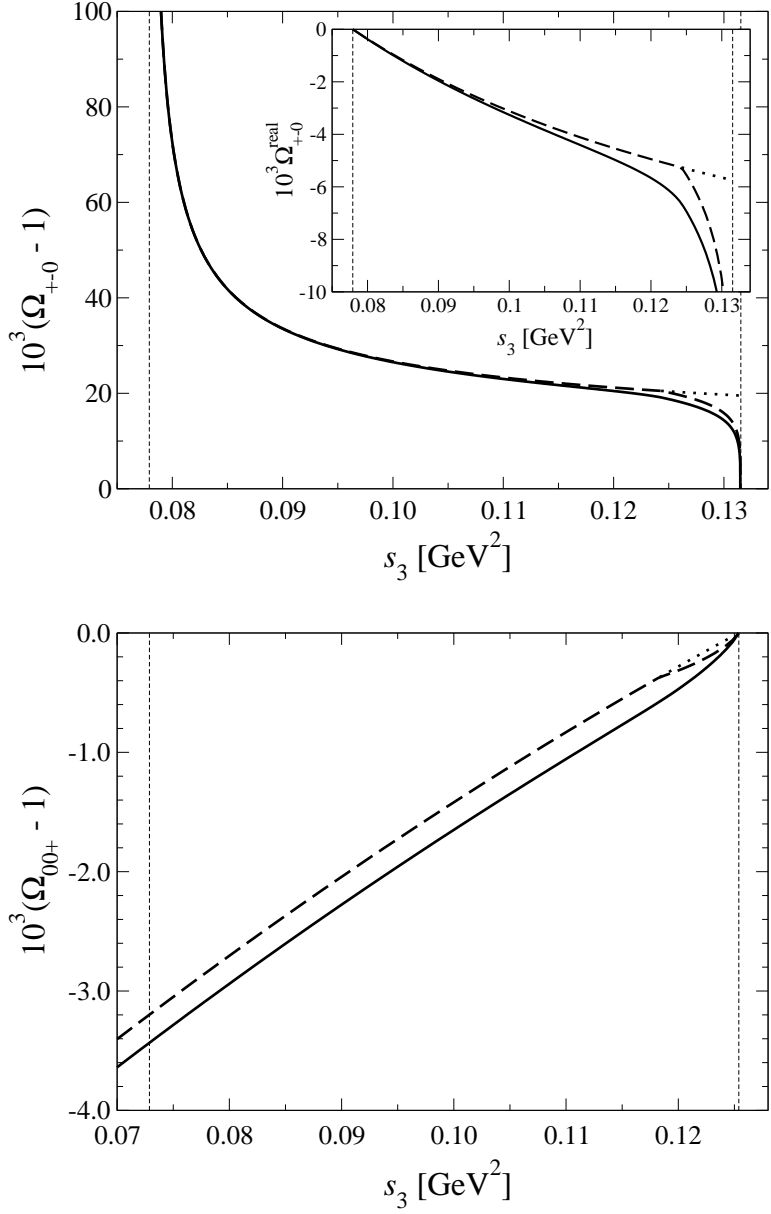

\begin{center}
\includegraphics[width=0.7\linewidth]{OmegaNexp.eps} \\[5mm]
\includegraphics[width=0.7\linewidth]{OmegaC.eps}
\end{center}
\caption{Comparison of the photon correction factors in the full calculation
(solid lines) and the soft-photon approximation (dashed/dotted lines), for 
the channels $K_L\to \pi^+\pi^-\pi^0$ (upper panel) and $K^+\to \pi^0\pi^0\pi^+$ (lower panel),
as functions of $s_3$.  We have employed a photon cutoff energy of $\Ec=10$~MeV.
In the upper panel, the main plot shows 
$\Omega_{+-0}(s_3,\Emaxn)$ (in units of $10^{-3}$), while
the insert represents 
$\Omega_{+-0}^{\rm real}(s_3,\Emaxn) \doteq \alpha/\pi\times \omega_{+-0}^{\rm real}(s_3,\Emaxn)$ alone,
i.e.\ with the Coulomb pole subtracted.
The lower panel shows $\Omega_{00+}(s_3,\Emaxp)$ (in units of $10^{-3}$).
The vertical lines denote the allowed kinematical range in $s_3$. 
For more details, see main text.}
\label{fig:softphoton}
\end{figure}
In Fig.~\ref{fig:softphoton}, these are plotted as functions of $s_3$ for
a fixed photon cut energy $\Ec=10$~MeV.
Full lines denote the complete calculation, dashed lines the soft-photon approximation.
In addition, the dotted lines show a ``naive'' soft-photon approximation with
$\Emaxp=\Ec$, i.e.\ one disregards the fact that the maximum photon energy
is, in some parts of phase space, not given by $\Ec$, but by the kinematical limit $\Ekin(s_3)$, 
see Eq.~\eqref{eq:Emax}.
In the other cases, the correction factors are not smooth at the point where
$\Ekin(s_3)=\Ec$. 

For the channel $K_L\to \pi^+\pi^-\pi^0$, we find that, as power counting suggests,
the Coulomb pole term in $\omega_{+-0}^{\rm virt}(s_3)$ completely dominates $\Omega_{+-0}$.
In the insert of the upper panel of Fig.~\ref{fig:softphoton}, this term has therefore
been subtracted to make the differences between the various curves more visible.
Even so, we find that the soft-photon approximation is a very small effect, 
and the difference in $\Omega_{+-0}$ stays well below $2\times 10^{-3}$.
However, there is a more significant deviation from the ``naive'' soft-photon 
approximation due to a logarithmic divergence at the upper kinematical limit for $s_3$, 
signalling that very close to the boundary of phase space, the approximation
of $\Order(\alpha)$ becomes unreliable.

For the channel $K^+\to \pi^0\pi^0\pi^+$, the correction factor is very small
throughout phase space ($|1-\Omega_{00+}|<4\times 10^{-3}$), and the deviation
due to the soft-photon approximation is smaller than $0.2\times 10^{-3}$.
Furthermore, except for the region where the maximum photon energy is restricted
by phase space, the correction factor is very close to linear and could therefore
be largely absorbed in a coupling constant shift of the polynomial part of the amplitude.

We therefore confirm for both these $K\to 3\pi$ decay channels that
a treatment of the bremsstrahlung corrections in the soft-photon approximation
leads to negligible modifications in the pertinent correction factors.

\subsection[$K^+\to\pi^+\pi^+\pi^-$]{\boldmath{$K^+\to\pi^+\pi^+\pi^-$}}

As the corrections to the soft-photon approximation for both $K_L\to \pi^+\pi^-\pi^0$
and $K^+\to \pi^+\pi^0\pi^0$ turn out to be negligible,
we confine ourselves to this approximation for the last channel
$K^+\to\pi^+\pi^+\pi^-$.
In this case, the universal radiative corrections cannot be formulated as a function
of the single variable $s_3$ any more, therefore we here define
$\Omega_{++-}(s_1,s_2,s_3,\Emaxp)$ in terms of the differential decay width
instead of the decay spectrum,
\begin{align}
d\Gamma \bigr|_{E_\gamma<\Emaxp} &= 
\Omega_{++-}(s_1,s_2,s_3,\Emaxp) d\Gamma^{\rm int}  ~, \label{eq:Omega++-}\\
\Omega_{++-}(s_1,s_2,s_3,\Emaxp) &= 1 + \frac{\alpha}{\pi} 
\Bigl( \omega_{++-}^{\rm virt}(s_1,s_2,s_3) + \omega_{++-}^{\rm real}(s_3,\Emaxp) \Bigr) + \Order(\alpha^2) 
~. \nonumber
\end{align}
We remark briefly
that the definition Eq.~\eqref{eq:Omega++-} 
(in contrast to Eqs.~\eqref{eq:Omega+-0}, \eqref{eq:defOmega00+})
is only unambiguous as long as the soft-photon approximation is applied,
which neglects the photon in overall momentum conservation.
Otherwise the allowed kinematic range for, say, $s_1$ and $s_3$ is different
for the radiative process as compared to the non-radiative one.

All diagrams shown in Fig.~\ref{fig:oneloop}a--f contribute for this channel,
and we find the following result for virtual-photon corrections:
\be
\mathrm{Re}\M_{++-} = \bigl\{ H_0 + H_1 (p_3^0-\mpc)\bigr\}
\biggl[ 1 + e^2 \biggl(  
\frac{1+\sigma_1^2}{16\sigma_1} + \frac{1+\sigma_2^2}{16\sigma_2} - \frac{1+\sigma_3^2}{16\sigma_3}
\biggr)\biggr] \,,
\ee
where we use the notation $\sigma_i = \sqrt{1-4\mpc^2/s_i}$.
This results in
\be
\omega_{++-}^{\rm virt}(s_1,s_2,s_3) = \frac{\pi^2}{2}
\biggl[ \frac{1+\sigma_1^2}{\sigma_1} + \frac{1+\sigma_2^2}{\sigma_2} 
- \frac{1+\sigma_3^2}{\sigma_3} \biggr] ~.
\ee
The soft-photon radiation contributions are calculated from the squared matrix element
\be
|\M_{++-}^\gamma|^2 = -e^2 H_0^2 \sum_{i,j=0}^3 Q_i Q_j \frac{p_ip_j}{(p_ik)(p_jk)} ~,
\ee
where we use $p_0=P$ and the charges $Q_0=Q_3=-1$, $Q_1=Q_2=1$ for compactness.
The significantly simpler calculation results in
\begin{align}
\omega_{++-}^{\rm real}(s_3,\Emaxp) &=
\frac{2\lambda_c}{3\mk^2\mpc^2} \Bigl(\ln\frac{2\Emaxp}{\mpc} - \frac{1}{3}\Bigr) 
+\Order(\e^4,\Emaxp)
~. 
\end{align}
The term $+\Order(\Emaxp)$ indicates once more that, 
in contrast to earlier sections, real photon emission is only treated in 
the soft-photon approximation.  

As for the other supplementary channel $K_L \to \pi^+ \pi^- \pi^0$,
we disregard radiative corrections of order $e^2a^1$, therefore
the amplitude $\M_{++-}^{\rm int}$ is given directly in terms of the 
amplitude $\M_{++-}$ in Ref.~\cite{CGKR,GKR,privComm}.

\section{On the accuracy of a determination of \boldmath{$a_0-a_2$}}\label{sec:accuracy}

In this section, we discuss the various sources of theoretical uncertainties
that limit the accuracy of an experimental determination of $a_0-a_2$ from
an analysis of the cusp in $K\to 3\pi$.  For definiteness, we concentrate
on the charged kaon decay channel $K^+ \to \pi^0 \pi^0 \pi^+$. 
The authors of Ref.~\cite{Cabibbo:2005ez} were the first to quote a
theoretical uncertainty on $a_0-a_2$, attributed to higher-order (three-loop)
terms in the expansion in $a$ as well as radiative corrections.  
A simple order-of-magnitude estimate of terms of $\Order(a^3)$ led 
to a combined uncertainty of 5\%.  This assessment was roughly confirmed
in Ref.~\cite{Gamiz}, where the uncertainty is quoted as 5--7\%. 

The analytic structure of the various $K\to 3\pi$ amplitudes
is now rather well understood, and combining the results of Refs.~\cite{CGKR,KlongLetter}
with the present investigation, one obtains a very accurate representation.
We identify the following main sources of theoretical uncertainty, in partial
agreement with Refs.~\cite{Cabibbo:2005ez,Gamiz}.
\begin{enumerate}
\item Electromagnetic corrections. \label{item:EM}
\item Isospin-breaking corrections in the matching of the coupling constants
$C_i$ to the $\pi\pi$ scattering lengths. \label{item:Matching}
\item Terms of order $a^2\e^4$. \label{item:ordera2e4}
\item Higher-order loop corrections in the strong sector, starting at 
$\Order(a^3 \e^3)$ in the non-relativistic power counting. \label{item:a3}
\end{enumerate}
Concerning point~\ref{item:EM}, we believe that the results presented in 
this article leave no significant error due to missing radiative corrections.
This clearly is the main progress compared to earlier investigations.
Concerning  point~\ref{item:Matching}, we have shown in Sect.~\ref{subsec:matching}
that the uncertainties  are of the order of one percent. Examples of terms at $\Order(a^2\e^4)$ 
mentioned in point~\ref{item:ordera2e4} can be worked out 
by comparing the exact form of the two-loop integral $F$ in Eq.~\eqref{eq:mostgeneral} 
with the explicit expressions at order $\e^2$,  provided in Ref.~\cite{KlongLetter},
or by partially adding effective range interaction vertices to the two-loop diagrams.

Finally, for an estimate of the uncertainty in point~\ref{item:a3}, we consider 
the \emph{threshold theorem}~\cite{fonda,Cabibbo:2004gq,CGKR}
for $K^+ \to \pi^0 \pi^0 \pi^+$ in the absence of photons: 
the leading non-analytic piece $\propto \sigma$
at the $\pi^+\pi^-$ threshold is given by the product of the decay amplitude
$K^+ \to \pi^+ \pi^+ \pi^-$ and the scattering amplitude $\pi^+\pi^- \to \pi^0\pi^0$, 
both evaluated at threshold. In particular, the coefficient of 
the cusp $\propto \sigma$ is proportional to 
\be
Z = T(K^+ \to \pi^+ \pi^+ \pi^-) \bigl|_{s_1 = 4\mpc^2,\,s_2=s_3=\frac{1}{2}(\mk^2-\mpc^2)} ~.
\label{eq:Zthreshold}
\ee
A calculation of the quantity $Z$ to $\Order(a^n)$ therefore yields the strength
of the leading cusp behavior as given in a representation of the decay amplitude
$K^+ \to \pi^0\pi^0\pi^+$ at $\Order(a^{n+1})$, in other words, the two-loop representation
of the $K^+ \to \pi^+ \pi^+ \pi^-$ amplitude given in Ref.~\cite{CGKR,GKR,privComm} allows for
an estimate of $a^3$ effects on the cusp.  
Numerically, the expansion of $Z$ up to two loops reads in arbitrary 
normalization\footnote{We use a rough estimate for the couplings
 $G_0,H_0$ worked out from  the slopes provided in PDG~\cite{pdg} and drop all derivative terms
proportional to the couplings $G_1$, $H_1$, \ldots, 
which generate small corrections only.}
\be
Z \,=\, -1.0\,(\textrm{tree}) \,-\, 0.13\,i \,(\textrm{1-loop}) 
\,+\, 0.014 \,(\textrm{2-loop}) ~, \label{eq:Znum}
\ee
i.e.\ terms of order $a^3$ will modify the strength of the leading (one-loop) cusp
by about 1.5\%.  The present estimate does not replace a full calculation of $\Order(a^3)$
effects, as it is necessarily incomplete: it neither yields a correct representation
of the amplitude elsewhere in the Dalitz plot (except in the cusp region), nor does
it give any information about subleading cusp behavior (e.g.\ $\propto \sigma^3$).  
However, we regard Eq.~\eqref{eq:Znum} as a good indication for the rate of 
convergence in the $K^+ \to \pi^0\pi^0\pi^+$ amplitude and the error in the latter
due to the omission of (strong) three-loop effects.

\medskip
We may conclude from the above that the decay amplitude for $K^+ \to \pi^0\pi^0\pi^+$ 
is indeed known very accurately.
However, the accuracy to which the amplitude is known does not necessarily translate
directly into the accuracy with which $a_0-a_2$ can be extracted from data.  
Rather, the latter is related to the change in the derivative of the amplitude
(or the decay spectrum) in the vicinity of the cusp. 
We illustrate this point with the help of the loop function $\JC(s_3)$, see
Eq.~\eqref{eq:replacementCoulomb}, which is responsible for the ``internal''
radiative corrections in $K^+ \to \pi^0\pi^0\pi^+$.  Below $\pi^+\pi^-$ threshold,
the singular part of it is given by
\be
-\frac{1}{16\pi} \bigl(\tilde\sigma + \alpha \log\tilde\sigma \bigr) ~, \quad
\tilde\sigma = \sqrt{\frac{4\mpc^2}{s_3}-1} ~.
\ee  
The derivative with respect to $s_3$ slightly below threshold is given by
\be
\frac{d}{ds_3}\biggl[
-\frac{1}{16\pi} \bigl(\tilde\sigma + \alpha \log\tilde\sigma \bigr)\biggr]_{s_3=4\mpc^2-\Delta}
= \frac{1}{64\pi\mpc\sqrt{\Delta}} \biggl( 1 + \frac{2\alpha\mpc}{\sqrt{\Delta}} \biggr) +\ldots ~,
\label{eq:JCderivative}
\ee
where higher-order terms in $\Delta/\mpc^2$ have been neglected.
So while the term of $\Order(\alpha)$ may represent a very small correction
on the absolute value of the amplitude, it changes the slope near the cusp, 
given by $\JC$ on a smooth (polynomial) background, by a relative factor of
$2\alpha\mpc/\sqrt{\Delta}$ that diverges near threshold.  For example, 
at the lower bound of the exclusion region in the NA48/2 experimental analysis~\cite{Batley},
$\Delta = \Delta_s = 5.25\times 10^{-4} \GeV^2$ (see Sect.~\ref{sec:pionium}), 
this relative factor is as large as 9\%.  

\medskip
We do not doubt that a more reliable estimate
of the theoretical uncertainty in $a_0-a_2$ is feasible
by working it out along with the data analysis.
 As all the
 ingredients are available now,  it would be useful to e.g.\ i) study the stability of
$a_0-a_2$ with respect to changes in the choice of the exclusion region $\Delta_s$,
as illustrated by the example above; ii) work out the influence of $a^2\e^4$ terms as described above;
iii) investigate the effect of Coulomb corrections at one vs.\ two-loop order.

\section{Comparison with other work}\label{sec:comparison}

Isidori~\cite{isidorirad} has considered the universal soft-photon corrections
in multi-body meson decays and applied this specifically to the decay channel
$K^+ \to \pi^+ \pi^+ \pi^-$.  
The results for the virtual photon corrections
correspond to a \emph{relativistic} calculation, 
as it is discussed in Appendix~\ref{app:loops}, i.e.\ they agree with
our non-relativistic results up to polynomial terms 
(that can be absorbed in a redefinition of the polynomial part of the decay amplitude)
and up to higher orders in the expansion in $\e$.  
Note that these higher-order terms in $\e$ are not complete in the sense that
there are non-universal corrections of the same order (some of which have 
been discussed as neglected higher-order diagrams in Sect.~\ref{sec:powercounting}).
The real photon radiation corrections are considered in the 
soft-photon approximation, the effects of which we have discussed in Sect.~\ref{sec:softphoton}.
Our results agree with those given in Ref.~\cite{isidorirad} (up to typos) when
expanding once more in $\e$.
Finally, Isidori~\cite{isidorirad}  mentions the formation of Coulomb bound states (such as pionium)
as one of the potentially relevant omissions in its discussion of radiative corrections.
We have closed this gap here in Sect.~\ref{sec:pionium}.

\medskip

In recent papers~\cite{tarasovk3piA,tarasovk3piB} the isospin-breaking corrections
to $K^\pm$ decays into three pions have been studied in a combined approach,
where the treatment of the cusps resembles that in the non-relativistic
effective field theory (see Refs.~\cite{CGKR,KlongLetter} and the present 
work), whereas the electromagnetic effects are included within a
quantum-mechanical approach. The treatment is not systematic. For example, 
in the strong sector the expression for the decay amplitude 
in our language corresponds to the summation of the one-loop
pion loops with non-derivative couplings only. This expression has not been
derived in Refs.~\cite{tarasovk3piA,tarasovk3piB} and, as follows from the present work, can be valid only 
in the vicinity of the cusp and only up to and including $\Order(\epsilon)$. 
The generalization to higher orders and beyond the cusp region is not 
discussed. The electromagnetic effects include the multi-Coulomb exchange
within the charged pion loop that corresponds roughly to our Eq.~\eqref{eq:replace}.
The procedure is again not systematic: Bremsstrahlung corrections, infrared divergences,
etc.\ are not discussed. We therefore conclude that the results of 
Refs.~\cite{tarasovk3piA,tarasovk3piB} cannot be safely used for the analysis of the 
experimental data on $K\to 3\pi$ decays.

\medskip 
The calculations of radiative corrections to $K\to 3\pi$ decays 
in Refs.~\cite{nehme,B1,B2}
are performed in the framework of Chiral Perturbation Theory.  
Such a framework is not suited to an extraction of the $\pi\pi$ scattering
lengths from experimental data, as these do not appear directly as parameters
of the theory.  
Furthermore, the effects corresponding to our corrections of $\Order(e^2 a)$
that modify the analytic structure of the cusp are beyond the accuracy
considered in Refs.~\cite{nehme,B1,B2} and are not considered there.

\section{Summary and conclusions}

In this article, we have generalized the framework of non-relativistic 
effective field theory for the cusp analysis in $K \to 3\pi$ decays
to include real and virtual photon effects.  
This framework now allows one to calculate the relevant amplitudes systematically
in a three-fold expansion in the non-relativistic expansion parameter $\e$, 
$\pi\pi$ scattering lengths $a$, and the electromagnetic coupling $e^2$.  
We have performed this calculation up to $\Order(e^2a^0\e^4)$ for all four 
$K\to 3\pi$ channels, and up to $\Order(e^2a^1\e^2)$ for those channels
in which the cusp effect is seen, namely $K^+ \to \pi^0\pi^0\pi^+$
and $K_L \to 3\pi^0$.
We have made consistency checks  with the corresponding relativistic
calculations of radiative corrections in various ways.
Real bremsstrahlung has been calculated without the soft-photon approximation
for $K^+ \to \pi^0\pi^0\pi^+$ and $K_L \to \pi^+\pi^-\pi^0$, and the effect of 
this approximation could explicitly be shown to be small.
The non-relativistic power counting provides a natural explanation for 
the common assumption that radiative corrections are dominated by Coulomb
photon exchange, and that e.g.\ bremsstrahlung effects are very small.
As an important phenomenon beyond the universal radiative corrections, 
we have studied the impact of pionium on the threshold (cusp) region.

We have provided simple analytic formulae for universal multiplicative correction
factors on decay spectra or differential decay widths, and for modified decay amplitudes
taking into account ``internal'' photon exchange processes.
We have commented on the various issues concerning the theoretical uncertainties
in the extraction of $a_0-a_2$ from the cusp effect.
We expect that using this formalism in a refined analysis of the experimental data
will lead to a precise determination of the $\pi\pi$ scattering lengths
from $K \to 3\pi$ decays.

\subsection*{Acknowledgments}

\begin{sloppypar}
We would like to thank the members of NA48/2 for many
useful discussions and a very fruitful collaboration, in particular
Brigitte Bloch-Devaux, Luigi Di Lella, Dmitry Madigozhin, and Italo Mannelli.
In particluar, we thank Luigi Di Lella and Dmitry Madigozhin 
for performing specific fits to $K\to 3\pi$ data  with our amplitudes. 
Furthermore, discussions with Sergey R.\ Gevorkyan and
Slawomir Wycech as well as useful e-mail exchanges 
with Gino Isidori are gratefully acknowledged.
Partial financial support under the EU Integrated Infrastructure
Initiative Hadron Physics Project (contract number RII3--CT--2004--506078)
and DFG (SFB/TR 16, ``Subnuclear Structure of Matter'') is gratefully
acknowledged. This work was  supported  by the Swiss
National Science Foundation, by EU MRTN--CT--2006--035482
(FLAVIA{\it net}), and by 
the Helmholtz Association through funds provided to the virtual 
institute ``Spin and strong QCD'' (VH-VI-231). 
 One of us (J.G.) is grateful to the Alexander von Humboldt--Stiftung and to 
the Helmholtz--Gemeinschaft for the award of  a prize 
 that allowed him to stay at the HISKP at the University of Bonn, 
where part of this work was performed. 
He also thanks the HISKP for the warm hospitality during these stays.
\end{sloppypar}

\clearpage

\renewcommand{\thefigure}{\thesection.\arabic{figure}}
\renewcommand{\thetable}{\thesection.\arabic{table}}
\renewcommand{\theequation}{\thesection.\arabic{equation}}

\appendix

\setcounter{equation}{0}
\setcounter{figure}{0}
\setcounter{table}{0}

\newpage

\section{Notation, kinematics}\label{app:notation}

In this appendix, we briefly collect some of the notation that is used throughout the article.
The masses of the charged pions are
denoted by $\mpc$ while for the neutral pion, $\mpn$ is used. 
Accordingly, $\mk$ denotes the charged kaon mass and $\mkl$ that of the $K_L$.
The decay channels are abbreviated by
indicating the charge of the final state pions, and the momenta are
assigned in the usual convention such that the odd pion carries four momentum $p_3$.

\begin{center}
\renewcommand{\arraystretch}{1.3}
\begin{tabular}{c|c}
decay channel & abbreviation \\ \hline
$K_L(P) \to \pi^0(p_1)\pi^0(p_2)\pi^0(p_3)$ & $\Knn$  \\
$K_L(P) \to \pi^+(p_1)\pi^-(p_2)\pi^0(p_3)$ & $\Knc$  \\ 
$K^+(P) \to \pi^0(p_1)\pi^0(p_2)\pi^+(p_3)$ & $\Kcn$  \\
$K^+(P) \to \pi^+(p_1)\pi^+(p_2)\pi^-(p_3)$ & $\Kcc$ 
\end{tabular}
\end{center}
The Mandelstam variables are defined as
\be
s_1 = (p_2+p_3)^2 \co \quad  \mbox{cycl.}
\ee
In the charged kaon rest frame, the energy and 3-momentum of the pions are given by
\eq
p_i^0=\frac{M_K^2+M_i^2-s_i}{2M_K}\, ,\qquad 
{\bf p}_i^2=\frac{\lambda(M_K^2,M_i^2,s_i)}{4M_K^2}
\en
(in the neutral kaon decays $M_K\to \mkl$). In the above equation,
we have used 
\begin{align}
\lambda(x,y,z) &= x^2+y^2+z^2-2xy-2xz-2yz\, .
\end{align}
Furthermore, the following functions of kinematical variables are frequently used:
\begin{align}
\lambda_0 &= \lambda\left(\mkl^2,s_3,\mpn^2\right)\co &
\lambda_c &= \lambda\left(\mk^2,s_3,\mpc^2\right)\co \nnnl
\sigma &= \sqrt{1-\frac{4  \mpc^2}{s_3}} \co &
q_0^2 &= \frac{s_3}{4}-\mpc^2 = \frac{s_3 \,\sigma^2}{4} \fs
\end{align}
The maximum photon energy $\Emaxp$ ($\Emaxn$) for $K^+$ ($K_L$) decays
in the kaon rest frame is given by
\begin{align}
\Emaxp^{(n)}  &= \left\{ \begin{array}{c@{\quad : \quad}l}
    \Ekin^{(n)} & \Ekin^{(n)} < \Ec \\
    \Ec & \Ekin^{(n)} > \Ec \end{array} \right. ~, \qquad \textrm{where}  \nnnl
\Ekin &= \frac{\mk^2-(\mpc+\sqrt{s_3})^2}{2\mk} ~,  \quad
\Ekinn = \frac{\mkl^2-(\mpn+\sqrt{s_3})^2}{2\mkl}\fs \label{eq:Emax} 
\end{align}

Loop integrations are denoted by an angle bracket and the pole in four
dimensions is contained in the term $\lambda$,
\begin{align}
\left\<\ldots\right> &=\int\frac{d^Dl}{i(2\pi)^D}\ldots \co  &D &= d+1 \co \nn
\lambda &= \frac{\mu^{D-4}}{16
  \pi^2}\left[\frac{1}{D-4}-\frac{1}{2}\left(\Gamma'(1)+\ln(4\pi)+1 \right)
\right]\fs \label{eq:app:lambda}
\end{align}
In the text, we also use -- depending on the context -- the symbols $\lambda_{IR}$ 
or $\lambda_{UV}$ for this quantity.

The differential decay width for the $n$-body decay of a particle of mass $M$
in $D=4$ dimensions is given in terms of the matrix element $\M$ according to
\begin{align}
d\Gamma &= N d\Phi_n(P;p_1,\ldots,p_n) |\M|^2 ~, \nnnl
N &= \frac{(2 \pi)^4}{2 M} \frac{1}{S} ~, \quad
d\Phi_n(P;p_1,\ldots,p_n) = \delta^4\Bigl(P-\sum_{i=1}^np_i\Bigr) 
\prod_{i=1}^n \frac{d^3 p_i}{2p_i^0(2\pi)^3} ~. \label{eq:PSelement}
\end{align}
where the values of the symmetry factor and the mass, $(S,M)$, are $(2,\mk)$ for
$\Kcn$ and $\Kcc$, $(1,\mkl)$ for $\Knc$ and $(6,\mkl)$ for $\Knn$.
For the phase space integrations in $D=d+1\neq 4$ dimensions, 
we generalize Eq.~\eqref{eq:PSelement} using
\begin{align}
& N(d) = \frac{(2 \pi)^D}{2 M} \frac{1}{S}\co \quad 
d\mu(p) = \frac{d^dp}{2p^0 (2\pi)^d}\co \quad 
\Omega_d =\frac{2\pi^\frac{d+1}{2}}{\Gamma(\frac{d+1}{2})} \co \nnnl
& d\Phi_n(P;p_1,\ldots,p_n,d) = \delta^D\Bigl(P-\sum_{i=1}^np_i\Bigr) 
\prod_{i=1}^n d\mu(p_i)  ~.
\end{align}

The cut of the logarithm is chosen, as usual, on the negative real axis and the $\psi$ function 
and the dilogarithm are defined as
\begin{align}
\psi(x) &= \frac{d}{dx}\ln
\Gamma(x) \co &\mbox{Li}_2(z) &= -\int_0^1 dx\, \frac{\ln(1-xz)}{x}\fs
\end{align}

\setcounter{equation}{0}
\setcounter{figure}{0}
\setcounter{table}{0}

\section{Radiative corrections: relativistic framework}
\label{app:radcorr}
The non-relativistic calculation presented in the main text is quite complex. In order to
have a fully independent check on that calculation, in this appendix we evaluate 
 radiative corrections  in a relativistic framework, for 
 the two decay channels $\Knc$ and $\Kcn$. 
In particular, we work out the corrections $\Omega_{+-0}$ and $\Omega_{00+}$ 
 (see Eqs.~\eqref{eq:Omega+-0}, \eqref{eq:defOmega00+}) in the relativistic framework 
and compare the
 result with the non-relativistic calculation described in the main text. 
 We confirm that, to the order considered in the momentum expansion,  the relativistic result
 indeed is identical to the non-relativistic one, up to a redefinition of tree-level couplings.
Note that, as the purpose of this Appendix is mainly
to cross-check the non-relativistic calculation, we also here disregard
all one-photon and one-pion reducible graphs (see e.g.\ Ref.~\cite{B1}),
which turn into polynomial terms when expanded non-relativistically in $\e$. 

In the present Appendix, the final results of this  calculation 
are presented. The pertinent contributions  include  virtual and real photons.
 The    evaluation of Bremsstrahlung  is performed 
in Appendix~\ref{app:phasespace}, whereas the relativistic loop integrals are 
detailed in Appendix~\ref{app:loops}. 

The calculation is performed in the Feynman-gauge. 
Infrared and ultraviolet divergences are both tamed with dimensional 
regularization. 
In keeping with the standard procedure for calculations
of radiative corrections in a relativistic framework, we retain the distinction
between both types of divergences in this appendix.
To make the formulae more transparent, 
we use the same scale for both, UV and IR divergences, although the origin of the 
divergences can be clearly identified. We therefore use the notation
\be
\lambda_{UV} = \frac{\mu^{D-4}}{16
  \pi^2}\left[\frac{1}{D-4}-\frac{1}{2}\left(\Gamma'(1)+\ln(4\pi)+1 \right)
\right]=\lambda_{IR}\,.
\ee
The loop integrals $\Lambda_+,\Lambda_0,T_i^{ab},\bar J^{ab}$ 
may be found in Appendix~\ref{app:loops}, 
and the Bremsstrahlung integrals $I_i^f$ 
are given in Appendix~\ref{app:phasespace}, together with the symbols $\delta_{0,c}$. 
As already mentioned in the main text, we evaluate real photon emission 
beyond the soft-photon approximation. 

\subsection[$K^+\to \pi^0\pi^0\pi^+$]{\boldmath{$K^+\to \pi^0\pi^0\pi^+$}}
We first consider the decay $K^+(P) \to \pi^0(p_1)\pi^0(p_2)\pi^+(p_3)$.
The relativistic interaction Lagrangian is 
\bea
\mathcal{L}_{\rm int} &=& \frac{G_0}{2} (1+e^2 b) ( K^\dagger \pi \pi^0  \pi^0+ \mathrm{h.c.}) \,\,,  \nn
b &=& 6 \lambda_{UV}+b_r(\mu) \,\, .
\eea
Here and in the following, we do not explicitly write the couplings to photons, because these are 
trivially  constructed via minimal coupling. 
\begin{figure}
\includegraphics[width=\linewidth]{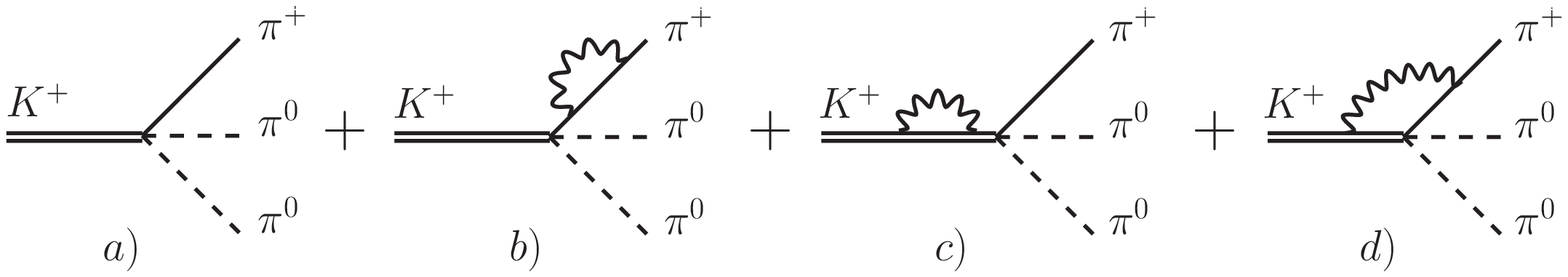}
\caption{Virtual photon corrections to the $K^+\to \pi^0\pi^0\pi^+$ decay.}
\label{fig:virtp00}
\end{figure}
The virtual corrections displayed in Fig.~\ref{fig:virtp00} generate the contribution
\be
\mathcal{M}^\mathrm{virt}(s_3) = G_0\left[Z - e^2\Lambda_+(s_3)\right] \,;\,s_3=(P-p_3)^2\,\fs\nonumber
\ee
The $Z$-factor is
\be \label{eq:zfactor}
Z = 1-4e^2\left(\lambda_{UV} - \lambda_{IR} \right) \fs
\ee
Since the amplitude depends only on $s_3$,
integration over phase space gives for the decay spectrum in $d$ dimensions
\bea\label{eq:phasespace}
\frac{d\Gamma}{ds_3} &=& N(d) (2\pi)^d \Phi_2(M_K^2,s_3,m_3^2,d) \Phi_2(s_3,m_1^2,m_2^2,d)
|\mathcal{M}^\mathrm{virt}(s_3)|^2 \co
\eea
where $\Phi_2$  denotes the two-particle phase space in $d$ dimensions, see
Eq.~\eqref{eq:2pps}, and where $p_i^2=m_i^2$.
We add Bremsstrahlung contributions, worked out in Appendix~\ref{app:phasespace}. 
These  cancel the infrared divergences generated by 
$\cal{M}^\mathrm{virt}$. Finally, we determine the relativistic analogon of the 
correction factor introduced in  Eq.~\eqref{eq:defOmega00+}. Using the 
notation $\Omega_{00+}\to\Omega^R_{00+}$, we find
\begin{align}
 \Omega^R_{00+}(s_3,\Emaxp) &=
 1+\frac{\alpha}{\pi}[F_1+F_2+F_3]+\Order(\alpha^2)\co \\
 F_1 &= A\, \left(L_E \, T_1^{K\pi}(s_3)-T_2^{K\pi}(s_3)\right) \co \nn
 F_2 &= \frac{\mk^2}{\sqrt{\lambda_c}}
 \left[2I_4^f(s_3,\Emaxp)-I_2^f(s_3,\Emaxp)-I_3^f(s_3,\Emaxp)  \right]
 \co\nn
 F_3 &= 8\pi^2b_r(\mu) - 3 \ln \frac{\mpc}{\mu} 
-2 L_E +\frac{\mpc^2}{\mk^2-\mpc^2} \ln \frac{\mk}{\mpc} -8\pi^2 \bar{J}^{K\pi}(s_3) \co \nonumber
\end{align}
where
\bea
L_E = \ln\left(\frac{2\Emaxp \mk}{\sqrt{\lambda_c}}\right) \co \,\,\,A
= \frac{\mk^2+\mpc^2-s_3}{\mpc^2}\fs
\eea

To compare with $\Omega_{00+}$, we expand  $\Omega^R_{00+}$ in $\e$ up 
to and including $\Order(\e^3)$, 
\begin{align}
\Omega^R_{00+}(s_3,\Emaxp)  &= 1+\frac{\alpha}{\pi}\biggl\{  
8 \pi^2 b_r(\mu) - 3 \ln \frac{\mpc}{\mu}  -\frac{5}{2} 
+\frac{3}{2}\Bigl(1-\frac{\mk\!-\!3\mpc}{2\mpc}\Bigr)\ln\frac{\mk}{\mpc} \nn
& + \frac{\lambda_c}{6\mk^2\mpc^2} \biggl[
\ln\frac{2\Emaxp}{\mpc} - \frac{97}{24} +\frac{19}{8}\ln 3
+ \frac{2}{3}\Bigl((4-\delta_c)\sqrt{1-\delta_c} -4\Bigr)
\nn & \qquad\qquad -2 \ln\frac{1+\sqrt{1-\delta_c}}{2} \biggr] +\Order(\e^4) \biggr\}
+\Order(\alpha^2) \fs
\end{align}
It is seen that $\Omega^R_{00+}$ differs from 
$\Omega_{00+}$ by a 
polynomial in the external momenta, which  can be eliminated by a redefinition of the 
tree-level coupling constants. 
We conclude that the relativistic calculation confirms the 
non-relativistic result for $\Omega_{00+}$ presented in the main text.

\subsection[$K_L\to \pi^+\pi^-\pi^0$]{\boldmath{$K_L\to \pi^+\pi^-\pi^0$}}\label{app:klc}
\begin{figure}
\includegraphics[width=\linewidth]{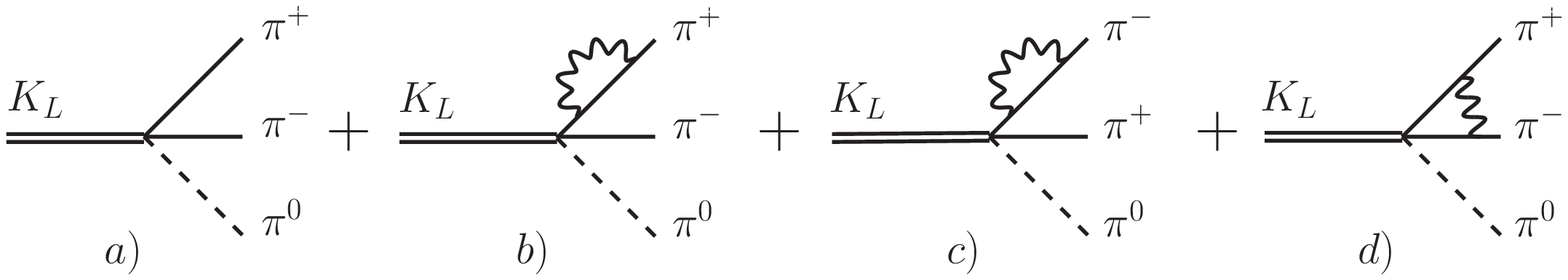}
\caption{Virtual photon corrections to the $K_L \to \pi^+\pi^-\pi^0$ decay.}\label{fig:virtKL}
\end{figure}
Here we consider the decay $K_L(P) \to \pi^+(p_1)\pi^-(p_2)\pi^0(p_3)$.
The relativistic interaction Lagrangian is 
\bea
\mathcal{L}_{\rm int} &=& L_0(1+e^2 a) K_L \pi^0 \pi^\dagger \pi \co  \nn
a &=& 6 \lambda_{UV} +a_r(\mu) \fs
\eea
The graphs displayed in Fig.~\ref{fig:virtKL} generate the amplitude
\begin{eqnarray}
\mathcal{M}^\mathrm{virt}(s_3) &=& L_0\left[Z+e^2\Lambda_0(s_3)\right]
 \, ;\, s_3=(P-p_3)^2\fs
\end{eqnarray}
The $Z$-factor was already  given in Eq.~\eqref{eq:zfactor}.
 We use the analogue of Eq.~\eqref{eq:phasespace},
and combine the result with  real photon emission considered in 
Appendix~\ref{app:phasespace}.  The relativistic correction factor
$\Omega^R_{+-0}$, the analogue of $\Omega_{+-0}$ in   Eq.~\eqref{eq:Omega+-0}, 
becomes
\begin{align}
\Omega^R_{+-0}(s_3,\Emaxn) &= 1+\frac{\alpha}{\pi} \left( f_1+f_2+f_3
\right)+ \Order(\alpha^2)\co\nn
f_1 &= 2B\, \left(\mbox{Re}\, T_2^{\pi\pi}(s_3)- \Big[1+L^n_E 
+\frac{1}{1+\sigma^2} \Big] \mbox{Re}\,T_1^{\pi\pi}(s_3)\right) \co \nn
f_2 &= -\frac{2\mkl^2 s_3}{\sqrt{\lambda_0} \mpc^2} \,\bigl[
  B\, \mbox{Re}\, T_1^{\pi\pi}(s_3) + 1\bigr] I_1^f(s_3,\Emaxn)  \co \nn
f_3 &= -\frac{5}{2}+8\pi^2a_r(\mu) - 3 \ln\frac{\mpc}{\mu}
-2L^n_E +\frac{\sigma}{2} \ln\Bigl(\frac{1+\sigma}{1-\sigma} \Bigr) \nn
& \quad  + \frac{1+\sigma^2}{2\sigma}
\left[ \mbox{Li}_2\Bigl(\frac{2 \sigma}{\sigma-1}\Bigr) - \mbox{Li}_2\Bigl(\frac{2
      \sigma}{\sigma+1}\Bigr)  \right] \co 
\end{align}
where
\begin{align}
L^n_E &=  \ln\left(\frac{2\Emaxn \mkl}{\sqrt{\lambda_0}}\right)\co  &B
&= \frac{1+\sigma^2}{1-\sigma^2} \fs
\end{align}
Expanding the result in $\e$ up to and including $\Order(\e^3)$, 
we obtain
\begin{align}
\Omega^R_{+-0} &= 1+\frac{\alpha}{\pi} \Bigg\{   \frac{\pi^2(1+\sigma^2)}{2 \sigma}
+ 8 \pi^2 a_r(\mu) - 3 \ln \frac{\mpc}{\mu}  -\frac{5}{2}  \nn
& \qquad + \frac{8\sigma^2}{3}
\bigg[\ln\frac{2 \Emaxn}{\mpc} -\frac{85}{24}
+2\sqrt{1-\delta_0} -2 \ln\Bigl(\frac{1+\sqrt{1-\delta_0}}{2}\Bigr) 
\bigg] +\Order(\e^4)  \Bigg\}  \nn & \quad+\Order(\alpha^2) ~.
\end{align}
This expression agrees with 
$\Omega_{+-0}$ in Eqs.~\eqref{eq:Omega+-0decomp}, \eqref{eq:omega+-0virt}, \eqref{eq:omega0} 
up to a polynomial in
the external momenta,  which can be absorbed with a redefinition of  the tree-level
couplings. We conclude that the relativistic calculation confirms the 
non-relativistic result for $\Omega_{+-0}$ presented in the main text.

\setcounter{equation}{0}
\setcounter{figure}{0}
\setcounter{table}{0}

\section{Bremsstrahlung beyond the soft-photon approximation}
\label{app:phasespace}
In this appendix, we evaluate the phase space integrals used in the
calculations of the bremsstrahlung corrections to $K \to 3\pi$ decays,
with a kinematical cut $\Ec$ in the photon energy in the rest frame of
the kaon. The evaluation of the decay spectra is, therefore, performed in
this frame. 

Most of the notation is set in Appendix~\ref{app:notation}.
In order to tame infrared singularities, we perform the
phase space integrations in $d$ dimensions and indicate this with an
additional argument $d$ in the infinitesimal $n$-body phase space
element $d\Phi_n$. The differential decay width is defined as
\bea
d\Gamma =N(d) d\Phi_4(P;p_1,p_2,q,k,d)\abs{\mathcal{M}}^2 \co
\eea
where $q$ and $k$ are the momenta of the odd pion and the photon, respectively.
In a first step, the four particle phase space is separated in a three particle phase space
integration over a two particle phase space integration,
\begin{align}
\int d\Phi_4(P;p_1,p_2,q,k,d) = (2\pi)^d \!\int ds_3\,
d\Phi_2(Q;p_1,p_2,d) d\Phi_3(P;Q,q,k,d) \co\nonumber
\end{align}
where $Q^2 = s_3$.

\subsection[$K^+ \to \pi^0\pi^0\pi^+$]{\boldmath{$K^+\to \pi^0\pi^0\pi^+$}}\label{app:brems00+}
We illustrate the calculation with the simplest possible
case $\Kcn$. The amplitude squared is given in Sect.~\ref{sec:kcn}.
After performing the two particle phase space integration, 
the integration over $Q$ and over the angle between the
photon and the charged pion as well as their angular parts, one is left with
the integrations over the photon- and the $\pi^+$-energy. The boundaries of
these integrations are given by
\begin{align}
q^0_\pm &=\frac{Y\pm k \sqrt{X}}{2\mk(\mk-2k)}\co &Y&=(\mk-k)(\mk^2+\mpc^2-s_3-2\mk k)\co \nn
X &=\lambda_c+4k\mk\Delta_+\co \!\! &\Delta_i&=s_3+M_{\pi^i}^2-\mk^2+\mk k\co\nn
\Ekin &= \frac{\mk^2-(\mpc+\sqrt{s_3})^2}{2\mk}\co&
\!\! \Emaxp &= \left\{ \begin{array}{c@{\quad : \quad}l}
    \Ekin & \Ekin < \Ec \\
    \Ec & \Ekin > \Ec \end{array} \right. \fs 
\end{align}
The integral reads
\bea\label{eq:dgkcn}
\frac{d\Gamma}{ds_3} &=& -G_0^2e^2(2\pi)^d N(d)
\Phi_2(s_3,\mpn^2,\mpn^2,d)\left[I_2+I_3-2I_4\right]\co\nn
I_i&=&\frac{\Omega_{d-2}\Omega_{d-1}}{8(2\pi)^{3d}}\int_0^{\Emaxp} dk
\,k^{d-5}\int_{q^0_{-}}^{q^0_{+}} dq^0 \,f_i(k,q^0,d)\co
\eea
where the two-particle phase space volume in $d$ dimensions is given by
\be \label{eq:2pps}
\hspace{-0.084cm}
\Phi_2(s,m_1^2,m_2^2,d) = \int d\Phi_2(P;p_1,p_2,d) = \frac{2}{(4
  \pi)^{\frac{3d}{2}}\Gamma(\frac{d}{2}) \sqrt{s}}
\bigg(\frac{\lambda(s,m_1^2,m_2^2)}{s}\bigg)^{\frac{d-2}{2}}
\ee
with $P^2 = s$, $p_i^2 = m_i^2\co$ and 
\begin{align}
f_2(q^0,k,d)&= (q \sqrt{1-z^{2}})^{d-3}\co &f_3(q^0,k,d) &=\mpc^2\frac{(q\sqrt{1-z^{2}})^{d-3}}{(q^0-q z)^2}\co\nn
f_4(q^0,k,d)&=q^0\frac{(q\sqrt{1-z^{2}})^{d-3}}{q^0-qz} \co &q &=
\sqrt{q_0^2-M_\pi^2} \co \nonumber
\end{align}
\be
 z =
\frac{M_K^2+\mpc^2-s_3+2(kq^0-M_Kq^0-kM_K)}{2k q}\fs \nonumber
\ee
We carry out the calculation in quite some detail for the integral $I_2$, the
other integrals can be performed along the same lines.

Introduce the variable $u$ as $q^0=\bar{q}^0+ku$
\bea
I_2&=&\frac{\Omega_{d-2}\Omega_{d-1}}{8(2\pi)^{3d}}\int_0^{\Emaxp} dk\,k^{d-4}\int_{u_{-}(k)}^{u_{+}(k)}du\,g(u,k)^{d-3}\co
\eea
with the functions
\begin{align}
\bar{q}^0&=\frac{\mk^2+\mpc^2-s_3}{2\mk}\co &\bar{q}&=\frac{\sqrt{\lambda_c}}{2\mk}\co & 
u_\pm(k) &=\frac{q^0_\pm-\bar{q}^0}{k}\co \nn
z&=\frac{\bar{q}^0-M_K(1+u)+ku}{q} \co&g(u,k) &= q \sqrt{1-z^{2}}\fs
\end{align}
To isolate the infrared singular part we
split up the function $g$ in a part with photon momentum zero and a remainder, 
\bea
g(u,k)^{d-3} &=& g(u,0)^{d-3}+\Delta g(u,k)\co\nn
\Delta g(u,k)&=&g(u,k)^{d-3}-g(u,0)^{d-3}\nn
&=&(d-3)\ln\left(\frac{g(u,k)}{g(u,0)}\right)+\Order\bigl((d-3)^2\bigr)\fs
\eea
Since the photon momentum dependent function $\Delta
g$ is proportional to $k^\alpha$ with $\alpha>0$, this integral vanishes
in the limit $d\to 3$ and we can drop it. The phase space integral has
an infrared divergence proportional to $(d-3)^{-1}$, therefore, in a next step,  
we expand the
remaining function $g(u,0)^{d-3}$ in the vicinity of $d=3$ and keep only
terms up to and including  $\Order(d-3)$.
Therefore, the original integral reads 
\begin{align}
I_2
&= \frac{\Omega_{d-2}\Omega_{d-1}}{8(2\pi)^{3d}}\mpc^{d-3}
\int_0^{\Emaxp}\!dk\,k^{d-4}\left[G_2(k)+(d-3)H_2(k)+\Order\bigl((d-3)^2\bigr)\right]
\co \nn
G_2(k)&=u_+(k)-u_-(k)\co \quad 
H_2(k) =\int_{u_{-}(k)}^{u_{+}(k)}du\,\ln\Bigl(\frac{g(u,0)}{\mpc}\Bigr)\fs\label{eq:decomposition}
\end{align}
Splitting up the functions $G_2(k)$ and $H_2(k)$ as well in a photon momentum
dependent and independent part, the original integral can be written as 
\be
I_2 = I_2^{IR}+\frac{1}{512\pi^7}I_2^f(s_3,\Emaxp)+\Order(d-3)\fs
\ee
where 
\begin{align}
I_2^{IR} &= G_2(0)\frac{\Omega_{d-2}\Omega_{d-1}}{8(2\pi)^{3d}}
\biggl(\frac{\Emaxp \mpc}{\mu^2}\biggr)^{d-3}\frac{\mu^{2(d-3)}}{d-3} \co\nn
I_2^f(s_3,\Emaxp) &=  H_2(0)+ \int_0^{\Emaxp} dk \frac{\Delta
      G_2(k)}{k} \co \nn
G_2(0) &= \frac{\sqrt{\lambda_c}}{\mk^2}\co \qquad H_2(0) =
\frac{\sqrt{\lambda_c}}{\mk^2}\left[\ln\left(\frac{\sqrt{\lambda_c}}{\mk \mpc}
  \right)-1\right] \co \nn
\frac{\Delta G_2(k)}{k} &= \frac{1}{\mk^2(\mk-2k)}\left[\frac{4\mk^2\Delta_+}{\sqrt{\lambda_c+4\mk k
    \Delta_+}+\sqrt{\lambda_c}} +2\sqrt{\lambda_c}\right]\fs 
\end{align}
The part $I_2^{IR}$ contains the infrared singularity of $I_2$ as well as
finite terms. Therefore, the part $I_2^f$ can be evaluated in $d = 3$.
Applying the same procedure to the remaining integrals, they divide
again into two parts, 
\be
I_i = I_i^{IR}+\frac{1}{512 \pi^7}I_i^f(s_3,\Emaxp)+\Order(d-3) \co
\quad i = 3,4 \co
\ee
and one finds
\begin{align}
I_i^f(s_3,\Emaxp) &= H_i(0)+ \int_0^{\Emaxp} dk \frac{\Delta
    G_i(k)}{k} \co\nn
I_3^{IR} &= I_2^{IR}\co \qquad I_4^{IR} = G_4(0) \frac{\Omega_{d-2}\Omega_{d-1}}{8(2\pi)^{3d}}
\left(\frac{\Emaxp\mpc}{\mu^2}\right)^{d-3} \frac{\mu^{2(d-3)}}{d-3}\co \nonumber
\end{align}
where
\begin{align}
\frac{\Delta G_3(k)}{k} &= \frac{4\Delta_+}{M_K( \sqrt{\lambda_c+4M_K k
    \Delta_+}+\sqrt{\lambda_c})}\co \nn
\frac{\Delta G_4(k)}{k} &= \frac{1}{M_K} \biggl[
\frac{\bar{q}^0}{k}\ln \left(\frac{1+D_+}{1+D_-}\right) 
- \ln \left(\frac{1+u_+(k)}{1+u_-(k)}\right) + u_+(k) - u_-(k) \biggr]
\co\nn
H_3(0) &= \frac{\sqrt{\lambda_c}}{2 \mk^2} \left[2 \ln\left(\frac{\sqrt{\lambda_c}}{\mk
      \mpc} \right)- \frac{ \bar{q}^0}{\bar{q}} \ln\left(\frac{\bar{q}^0+\bar{q}}{\bar{q}^0-\bar{q}} \right) \right]\co\nn
 H_4(0) &= \frac{\bar{q}^0}{2 \mk} \left[2 \ln \left(\frac{\sqrt{\lambda_c}}{\mk \mpc}
  \right) \ln \left(\frac{\bar{q}^0+\bar{q}}{\bar{q}^0-\bar{q}} \right)
  +\mathrm{Li}_2 \left(\frac{2\bar{q}}{\bar{q}-\bar{q}^0} \right) -
  \mathrm{Li}_2 \left(\frac{2\bar{q}}{\bar{q}+\bar{q}^0} \right) \right] , \nn
G_4(0) &= \frac{\bar{q}^0}{M_K}\ln\left(\frac{1+u_+(0)}{1+u_-(0)}\right)\co
\nn
D_\pm &= \pm \frac{\sqrt{\lambda_c+4M_K k
    \Delta_+}-\sqrt{\lambda_c}\mp 2 k M_K}{2 M_K \bar{q}^0\pm\sqrt{\lambda_c}} \fs
\end{align}
The remaining integrations over $u$ and $k$ can be done numerically in three
dimensions, as they are free of infrared divergences.

The bremsstrahlung decay spectrum  can be simplified
considerably by expanding the infrared finite part of the combination $I_2+I_3-2I_4$
that appears in Eq.~\eqref{eq:dgkcn} in the non-relativistic
parameter $\epsilon$. We find
\begin{align}
H_2(0) &+ H_3(0) -2H_4(0) 
~=~
\frac{\lambda_c^{3/2}}{36\mk^4\mpc^2}\biggl(5-6\ln\frac{\sqrt{\lambda_c}}{\mk\mpc}\biggr)
+\Order(\epsilon^5) \co \nonumber \\[1ex]
\int_0^{\Emaxp} dk & \frac{\Delta G_2(k)+\Delta G_3(k)-2\Delta G_4(k)}{k}  \nn &=
\frac{\lambda_c^{3/2}}{9\mk^4\mpc^2}
  \biggl(4 - (4-\delta_c)\sqrt{1-\delta_c} +3
  \ln\frac{1+\sqrt{1-\delta_c}}{2} \biggr) +\Order(\epsilon^5)\co
  \nonumber \\[1ex]
\delta_c &= \frac{8\mk(\mk-\mpc)\mpc\Emaxp}{\lambda_c} \fs
\end{align}

\subsection[$K_L \to \pi^+\pi^-\pi^0$]{\boldmath$K_L \to \pi^+\pi^-\pi^0$}\label{app:brems+-0}
In the decay channel $\Knc$, the four particle phase space is again separated
in a three particle phase space integration over a two particle phase space
integration. However, because the amplitude squared indicated in Sect.~\ref{sec:knc}
depends on the momenta of the two particle phase space integration, this two
particle phase space does not factorize anymore, but produces the functions
$R_1$ and $R_2$,
\begin{align}\label{eq:dgknc}
\frac{d\Gamma}{ds_3}&= 
-8 L_0^2 e^2 N(d) \Omega_{d-2} \Phi_2(s_3,\mpc^2,\mpc^2,d)\Gamma\left(\frac{d}{2} \right) \frac{(4\pi)^\frac{d}{2}}{1-\sigma^2} \left[R_1 - \frac{1+\sigma^2}{2} R_2 \right] I_1 \co\nn
R_1 &= 1+(d-3)  \left( \ln 2 -
\frac{1}{2\sigma}L \right) \co \qquad L = \ln\frac{1+\sigma}{1-\sigma}\co \nn
R_2 &= \frac{1}{\sigma}L  +  \frac{d-3}{2\sigma} 
\left(2\ln 2 \,L  + \Li{\frac{2\sigma}{\sigma-1}} - \Li{\frac{2\sigma}{\sigma+1}}  \right)
  \co
\end{align}
where $R_1$ and $R_2$ are evaluated in a vicinity of $d = 3$ and the
two particle phase space volume
$\Phi_2$ is given in Eq.~\eqref{eq:2pps}.
The remaining three particle phase space integrals can be solved along the lines
of the previous section, and one obtains
\bea
I_1 &=& I_1^{IR}+\frac{1}{512\pi^7}I_1^f(s_3,\Emaxn)\co\nn
I_1^{IR} &=& G_1(0) \frac{\Omega_{d-2}\Omega_{d-1}}{8(2\pi)^{3d}}
\left(\frac{\Emaxn\mpc}{\mu^2}\right)^{d-3} \frac{\mu^{2(d-3)}}{d-3} \co \nn
I_1^f(s_3,\Emaxn) &=&   H_1(0) + \int_0^{\Emaxn} dk \frac{\Delta
    G_1(k)}{k} \co \nn
\frac{\Delta G_1(k)}{k} &=& \frac{4 \Delta_0 \mpc^2}{\mkl s_3 \left(\sqrt{\lambda_0+4\mkl
      k \Delta_0} + \sqrt{\lambda_0} \right)}\co\nn
H_1(0) &=& \frac{\mpc^2 \sqrt{\lambda_0}}{\mkl^2 s_3}
\left[\ln\left(\frac{\sqrt{\lambda_0}}{\mkl \mpc} \right)-
  \frac{\bar{Q}_0}{2\bar{Q}}
  \ln\left(\frac{\bar{Q}_0+\bar{Q}}{\bar{Q}_0-\bar{Q}} \right) \right] \co\nn
G_1(0) &=& \frac{\mpc^2 \sqrt{\lambda_0}  }{\mkl^2 s_3}\co
\eea
where
\begin{align}
\bar{Q}_0 &= \frac{\mkl^2+s_3-\mpn^2}{2 \mkl}\co &\bar{Q} &=
\frac{\sqrt{\lambda_0}}{2 \mkl} \fs
\end{align}
Since the kinematics in the channel $\Knc$ differs from the one in $\Kcn$, the
integration boundary has to be changed to
\begin{align} 
\Ekinn &= \frac{\mkl^2-(\mpn+\sqrt{s_3})^2}{2\mkl}\co&
\Emaxn &= \left\{ \begin{array}{c@{\quad : \quad}l}
    \Ekinn & \Ekinn < \Ec \\
    \Ec & \Ekinn > \Ec \end{array} \right. \fs
\end{align}
Again, expanding the infrared finite part of $I_1$ in $\epsilon$ leads to
substantial simplifications,
\bea
H_1(0) &=& \frac{\sqrt{\lambda_0}}{4\mkl^2}
            \biggl( \ln\frac{\sqrt{\lambda_0}}{\mkl\mpc} -1 \biggr) +\Order(\epsilon^3)~,\nn
\int_0^{\Emaxn} dk \frac{\Delta G_1(k)}{k} &=& \frac{\sqrt{\lambda_0}}{2\mkl^2}
  \biggl( \sqrt{1-\delta_0}-1-\ln\frac{1+\sqrt{1-\delta_0}}{2} \biggr) +\Order(\epsilon^3)~,\nn
\delta_0 &=& \frac{8\mkl(\mkl-\mpn)\mpn\Emaxn}{\lambda_0} ~.
\eea
Furthermore, we find for the combination appearing in Eq.~\eqref{eq:dgknc}
\be
\frac{1}{1-\sigma^2} \left[R_1 - \frac{1+\sigma^2}{2} R_2 \right]
= -\frac{4}{3}\sigma^2 \biggl[1+\Bigl(\ln 2-\frac{5}{6}\Bigr)(d-3)\biggr]
+ \Order(\epsilon^4) ~.
\ee
Note that $\lambda_0=\Order(\epsilon^2)$ and $\sigma = \Order(\epsilon)$.

\subsection{Soft-photon approximation}\label{app:softphotonanalytic}

Performing the soft-photon approximation  in the four particle phase space
integration amounts, in the present case, to discard the photon 
momentum in the $\delta$-function,
\bea
d\Gamma =N(d)d\Phi_3(P;p_1,p_2,q,d) d\mu(k) \abs{\mathcal{M}}^2\fs
\eea
Note that energy and momentum are not conserved anymore.
Comparing the exact result calculated in the previous Sects.~\ref{app:brems00+}, \ref{app:brems+-0}
with the soft-photon approximation shows that
\begin{itemize}
\item[i)] the soft-photon approximation reproduces $I_i^{IR}$ in a vicinity of
  $d = 3$ 
\item[ii)] the difference between the two results is analytic in $\Emaxp$ and
  of  $\Order(\Emaxp)$.
\end{itemize}
The soft-photon approximation  $I_i^{S}$ for the
integral $I_i$ can be written as
\be
I_i^{S} = I_i^{IR}+\frac{1}{512\pi^7} I_i^f(s_3,0)+\Order(d-3)\co \quad i
= 1,\ldots,4\fs
\ee
For a discussion of the numerical size of the error induced by the soft-photon 
approximation in the non-relativistic limit, see Sect.~\ref{sec:softphoton}.

\vskip1cm

\setcounter{equation}{0}
\setcounter{figure}{0}
\setcounter{table}{0}

\section{Relativistic loop integrals}\label{app:loops}
Here we provide the loop integrals needed in the relativistic calculation presented 
in Appendix~\ref{app:radcorr}.  In particular,
 we evaluate the vertex functions $\Lambda_0$ and $\Lambda_+$, 
generated by the vertex graphs displayed in Fig.~\ref{fig:virtKL} 
and Fig.~\ref{fig:virtp00}, respectively. 
 In this appendix, pion and kaon momenta refer to charged particles throughout,
 $p_i^2=M_\pi^2$, $P^2=M_K^2$.
\begin{align}
\Lambda_0(s)&=\left\<\frac{g^{\mu\nu}}{-l^2}\frac{(2p_1-l)_\mu}{\mpc^2-(p_1-l)^2}\frac{(2p_2+l)_\nu}{\mpc^2-(p_2+l)^2}\right\>
\nn[2mm] 
&=2 (s-2\mpc^2)
  G(s,\mpc^2,\mpc^2)-J^{\pi\pi}(s)+2 J^{\pi\gamma}(\mpc^2) \co\nn[2mm]
\Lambda_+(t)&= \left\<\frac{g^{\mu\nu}}{-l^2}\frac{(2P-l)_\mu}{\mk^2-(P-l)^2}\frac{(2p_3-l)_\nu}{\mpc^2-(p_3-l)^2}\right\>  \nn[2mm] 
&= 2 (\mk^2+\mpc^2-t)
  G(t,\!\mk^2,\!\mpc^2)+J^{K\pi}(t)  -J^{\pi\gamma}(\mpc^2)-J^{K\gamma}(\mk^2) \co \nn[2mm]
J^{ab}(p^2) &= \left\< \frac{1}{M_a^2-l^2}\frac{1}{M_b^2-(p-l)^2}
\right\>\co \nn[2mm]
G(s,\! M_a^2,\!M_b^2) &= 
\left\<\frac{1}{(-l^2)(M_a^2-(p_1-l)^2)(M_b^2-(p_2+l)^2)}\right\> \co
\end{align}
where $s=(p_1+p_2)^2$, $t=(P-p_3)^2$.
The evaluation of the integrals $J^{ab}$ is standard.
In Appendix~\ref{app:radcorr}, we use the notation
$ \bar{J}^{ab}(p^2) = J^{ab}(p^2)-J^{ab}(0) $.
The vertex integral is
\begin{align}
G(s,M_a^2,M_b^2) &= 
2\int_0^1dxydy\left\<\left[-l^2+y^2 M_b^2\tau(x)\right]^{-3}\right\>\nn
&= \frac{M_b^{D-6}}{D-4}\frac{1}{(4\pi)^{\frac{D}{2}}}\Gamma\Bigl(3-\frac{D}{2}\Bigr)
\int_0^1 dx\, \tau(x)^{\frac{D}{2}-3}\,\,,\nn
M_b^2 \tau(x) &= M_b^2(1-x)+ M_a^2 x- s x(1-x)\,\,.\nonumber
\end{align}
Expanding around  $D = 4$, we find
\be
G(s,M_a^2,M_b^2) = \frac{1}{M_b^2}
\biggl\{\lambda_{IR} + \frac{1}{16\pi^2}\Bigl(\ln\frac{M_b}{\mu}+\frac{1}{2}\Bigr)\biggr\}T_1^{ab}(s)
+\frac{T_2^{ab}(s)}{16\pi^2 M_b^2} + \Order(D-4)\co \label{eq:Gs}
\ee
with
\begin{align}
\left\{T_1^{ab};T_2^{ab}\right\}=\frac{1}{2}\int_0^1\frac{dx}{\tau(x)}\left\{2;\ln(\tau(x))\right\}\fs
\end{align}
For equal masses $M_a = M_b = M_\pi$, we find
\begin{align}
T_1 &= -\frac{1-\sigma^2}{2\sigma} \left[\ln\left( \frac{1+\sigma}{1-\sigma}\right)- i \pi \right] \co \nn
T_2 &= \frac{1-\sigma^2}{4\sigma} \bigg[\pi^2 +\Li{\frac{2\sigma}{\sigma-1}}
-\Li{\frac{2\sigma}{\sigma+1}}
 + \pi i
\ln\Bigl(\frac{4\sigma^2}{1-\sigma^2} \Bigr) \bigg] \co \nn
 \sigma &= \sqrt{1-\frac{4\mpc^2}{s}}\co \quad s\geq 4M_\pi^2\fs 
\end{align}

\setcounter{equation}{0}
\setcounter{figure}{0}
\setcounter{table}{0}

\section{Infrared divergences}\label{app:infra-ultra}

Throughout this paper, we have used dimensional regularization
to tame both ultraviolet and infrared divergences. In dimensional 
regularization, applying the threshold expansion,
one sets so-called no-scale integrals to zero.
While this simplifies calculations considerably, the identification of 
infrared singularities can be complicated. We shall demonstrate this explicitly
on the basis of the example of the vertex diagram, given in Figs.~\ref{fig:demo}b+c.

Let us start from the relativistic case. The pertinent scalar integral is
given by the function $G(s) \doteq G(s,\mpc^2,\mpc^2)$ discussed in 
Appendix~\ref{app:loops}.
As shown there, the integral is ultraviolet-finite.
We can rewrite Eq.~\eqref{eq:Gs} according to
\eq\label{eq:Grel}
{\rm Re}\,G(s)&=&-\frac{2}{s\sigma}\ln \left(\frac{1+\sigma}{1-\sigma}\right) 
\lambda_{IR}+{\rm Re}\,G^r(s)
\nonumber\\[2mm]
&=&-\frac{1}{M_\pi^2}\,\lambda_{IR}+\Order({\bf q}^2)
+{\rm Re}\,G^r(s)\, ,
\en
where ${\bf q}$ is the relative momentum of the pion pair in the final state,
$\sigma=\sqrt{1-4M_\pi^2/s}$ and the quantity 
${\rm Re}\,G^r(s)$ is finite as $D\to 4$.
The quantity $\lambda_{IR}$ is defined by Eq.~\eqref{eq:app:lambda}. 
In order to emphasize that the divergence at $D\to 4$ is infrared, the 
subscript ``IR'' is attached.

Let us now consider the same diagram in the non-relativistic effective theory.
To this end, one may replace the pion propagators
\eq
\frac{1}{M_\pi^2-p^2}\to
\frac{1}{2M_\pi}\,\frac{1}{M_\pi+{\bf p}^2/(2M_\pi)-p^0}
+\ldots
\en
(for simplicity, we use here the standard version of the non-relativistic
EFT). Performing the contour integration over $l^0$, at lowest order
in the momentum expansion we arrive at
\eq
G^{NR}(s)=\frac{1}{4M_\pi}\,\int\frac{d^dl}{(2\pi)^d}\,
\frac{1}{{\bf l}^2}\,\biggl(\frac{1}{{\bf l}^2-2{\bf q}{\bf l}}
-\frac{1}{2M_\pi}\,\frac{1}{|{\bf l}|-{\bf q}{\bf l}/M_\pi+{\bf l}^2/(2M_\pi)}
\biggr) .~
\en
The real part of the first term is finite as $D\to 4$.
Consequently,
\eq
G^{NR}(s)=-\frac{1}{8M_\pi^2}\,\int\frac{d^dl}{(2\pi)^d}\,
\frac{1}{{\bf l}^2}\,\frac{1}{|{\bf l}|+{\bf l}^2/(2M_\pi)}
+\Order({\bf q}^2)+\mbox{finite at }D\to 4\, .
\en
Performing the remaining integration, we exactly reproduce the infrared divergence
in Eq.~\eqref{eq:Grel}. However, if the integrand is first expanded in 
inverse powers of $M_\pi$, each term in this expansion is a no-scale 
integral and vanishes in dimensional regularization. Thus, 
if threshold expansion is applied, the non-relativistic EFT fails to reproduce
the infrared divergences of the relativistic theory.

In order to understand this apparent contradiction, note that only the first term in
the threshold expansion, which is singular at the origin as $|{\bf l}|^{-3}$,
 is infrared-divergent. Introducing a splitting of the integration interval, 
one finds
\eq
&&\frac{1}{8M_\pi^2}\,\lim_{A\to 0,C\to \infty}
\biggl(\int_A^B+\int_B^C\biggr)\frac{d^dl}{(2\pi)^d}\,\frac{1}{|{\bf l}|^3}
\nonumber\\[2mm]
&=&\frac{1}{M_\pi^2 2^{d+2}\pi^{d/2}\Gamma(\frac{d}{2})}\,\lim_{A\to 0,C\to \infty}
\biggl(\frac{1}{d-3}\,(B^{d-3}-A^{d-3})+\frac{1}{d-3}\,(C^{d-3}-B^{d-3})
\biggr)
\nonumber\\[2mm]
&=&\frac{1}{M_\pi^2}\,(\lambda_{IR}-\lambda_{UV}) + \Order(d-3)\, .
\en
In the second line, we have taken $d>3$ ($d<3$) in the first (second) term. 
We conclude that, although the above expression is formally zero, one may 
identify parts of this expression with infrared and ultraviolet 
divergences. The infrared divergences, which are present in the relativistic
theory, are reproduced in non-relativistic EFT.

To summarize, using no-scale arguments in  dimensional regularization changes
the structure of infrared divergences in the non-relativistic EFT, 
since it amounts to setting
expressions of the type  $\lambda_{IR}-\lambda_{UV}$ to zero. 
It is further  seen that the change in the amplitude is a low-energy polynomial, because
of the presence $\lambda_{UV}$ in the ``dropped'' term. Then, owing to 
the fact that the infrared divergences cancel at the end, one may justify
using the shortcut, based on threshold expansion and removing all divergences
at $D\to 4$ by the counterterms in the Lagrangian.

\setcounter{equation}{0}
\setcounter{figure}{0}
\setcounter{table}{0}

\section{Phase space in  \boldmath{$K_L\to 3\pi^0$}}\label{app:dalitzkl3pi0}
\begin{figure}
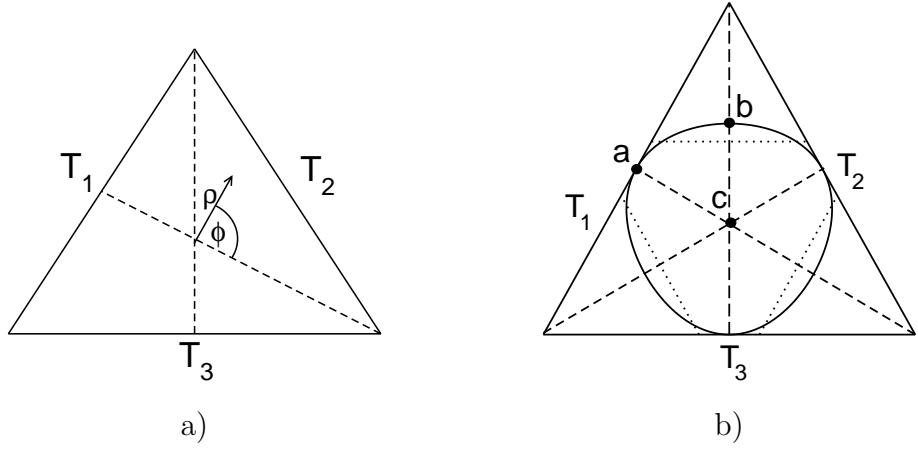

\begin{center}
\begin{tabular}{cccccc}
\includegraphics[bb= 122 21 673 541, width=5.cm]{coordinates.eps}\hspace{2cm}&
&&&&\includegraphics[bb = 161 32 634 512,width=5cm]{dalitz_kl3pi.eps}\\[2mm]
a)&&&&&b)\\[2mm]
\end{tabular}
\caption{The Dalitz-plot for $K_L\to 3\pi^0$ decay. In Fig.~a), we display the polar coordinates,
with $\rho \doteq Q\,r/3$.  
The physical region for the decay is bounded by the curve  displayed in Eq.~\eqref{eq:curve_r}, shown
 in b) with a solid line. The three dotted lines denote the location of the cusps at $s_i=4M_\pi^2$.
  By a proper choice of the outgoing momenta of the pions, all events may be mapped into 
the region $a b c$.}
\label{fig:dalitz}
\end{center}
\end{figure}

\begin{figure}
\begin{center}
\includegraphics[width=9.cm]{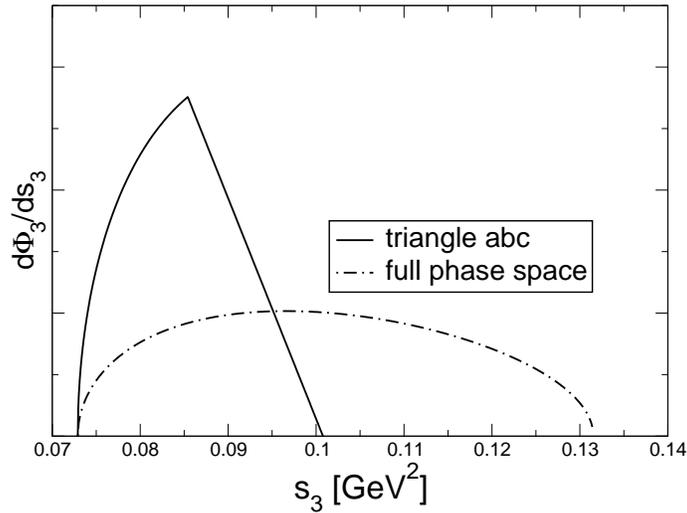}
\caption{The differential phase space $d\Phi_3/ds_3$. Its shape depends on the region chosen. The dash-dotted line is the standard full phase space region. The solid line corresponds 
to the case where all events are mapped into the triangle $abc$, see Fig.~\ref{fig:dalitz}b).}
\label{fig:phasespace}
\end{center}
\end{figure}

In case of the decay $K_L\to 3\pi^0$, it is  
convenient  to take advantage of the symmetry of the final state which contains 
3 identical particles~\cite{dalitz,fabri}. One introduces polar 
coordinates for the kinetic energies $T_i = p_i^0-\mpn$
according to
\bea
T_1&=&\frac{Q}{3}\left (1+r\cos{\phi}\right)\co\nonumber\\
T_2&=&\frac{Q}{3}\left\{1+r\cos{\Bigl(\phi+\frac{2\pi}{3}\Bigr)}\right\}\co\nonumber\\
T_3&=&\frac{Q}{3}\left\{1+r\cos{\Bigl(\phi-\frac{2\pi}{3}\Bigr)}\right\}\co
\eea
see Fig.~\ref{fig:dalitz}a for the definition of the variables $r,\phi$. In  Fig.~\ref{fig:dalitz}b, we show
 the boundary of the physical region
\bea\label{eq:curve_r}
(1+x)r^2+xr^3\cos{3\phi}=1\,,
\eea
where $Q=M_{K_L}-3M_{\pi^0}$ is the energy release, and 
$x=2\epsilon/(2-\epsilon)^2, \epsilon=Q/M_{K_L}$.
The differential decay width is 
\bea
d\Gamma=\frac{(2\pi)^4}{3!2M_{K_L}}d\Phi_3(P;p_1,p_2,p_3)|{\cal{M}}|^2\,.
\eea
Because the matrix element ${\cal M}$ is a Lorentz scalar, one can immediately 
perform seven of the nine 
integrations  over the pion momenta, without a need to know the matrix element ${\cal M}$. 
 Depending on the variables chosen, one has
\be
\int (2\pi)^9d\Phi_3 = 
\int C_{23}\,ds_2\,ds_3 = 
\int C_{r\phi}\,r\,dr\,d\phi = 
\int C_{s\phi}\,ds_3\,d\phi\,,
\ee
with Jacobi determinants 
\bea
C_{23}=\frac{\pi^2}{4M_{K_L}^2}\scs
C_{r\phi}=\frac{\pi^2\sqrt{3}}{18}Q^2\scs
C_{s\phi}=\frac{\pi^2\sqrt{3}}{24M_{K_L}^2}\frac{|M_{K_L}^2+3M_{\pi^0}^2-3s_3|}
{\cos^2{(\phi-\frac{2\pi}{3})}}\,.
\eea
 We note that the momenta of the 
three outgoing pions can always be chosen such that the decay products are located in the 
triangle $a b c$ in Fig.~\ref{fig:dalitz}b. This is a perfectly legitimate procedure, because the matrix element is totally symmetric in the variables $s_1,s_2$ and $s_3$ in this channel. 
In case that the integration is restricted to the triangle $ a b c$, the 
pertinent Jacobi-determinant must be multiplied with $3!$, because there are six identical 
contributions to the decay probability.
In Fig.~\ref{fig:phasespace}, we show the differential phase space in the standard case where 
the full phase space is used, and in the case where the events are mapped into 
the triangle $a b c$. Although the integrated phase space is of course identical in
the two cases, their shape is very different.
In the main text, we use the variables $s_3$, $\phi$ and restrict the phase space to the triangle $abc$.
Note that by this procedure, the number of neutral 
pion pairs located in the vicinity of the cusp region is increased by a factor of 3.

\end{document}